\documentclass[12pt]{article}
\usepackage{amsmath,amssymb,amsfonts}
\usepackage{a4wide,epsfig,psfrag,scalefnt}
\usepackage[dvipsnames]{xcolor}
\usepackage{braket}
\usepackage{placeins}
\usepackage{breqn}
\usepackage{slashed}
\usepackage{enumitem}
\usepackage[numbers,sort&compress]{natbib}
\usepackage{caption}
\usepackage{subcaption}
\usepackage[export]{adjustbox}

\usepackage[pdftitle={Nonleptonic B-meson decays to next-to-next-to-leading order},
  pdfauthor={Manuel Egner, Matteo Fael, Kay Sch\"onwald, and Matthias Steinhauser},
  pdfkeywords={},
  bookmarks=true, linktocpage,
  colorlinks=true, allbordercolors=white, allcolors=blue]{hyperref}

\graphicspath{ {figs/} }

\newcommand{\lr}{l_\rho}

\allowdisplaybreaks

\begin{document}

\title{\vskip-3cm{\baselineskip14pt
    \begin{flushleft}
     \normalsize CERN-TH-2024-089, P3H-24-041, TTP24-020, ZU-TH 30/24
    \end{flushleft}} \vskip1.5cm
  Nonleptonic $B$-meson decays to next-to-next-to-leading order
  }

\author{
  Manuel Egner$^{a}$,
  Matteo Fael$^{b}$,
  Kay Sch\"onwald$^{c}$,
  Matthias Steinhauser$^{a}$
  \\
  {\small\it (a) Institut f{\"u}r Theoretische Teilchenphysik,
    Karlsruhe Institute of Technology (KIT),}\\
  {\small\it Wolfgang-Gaede Stra\ss{}e 1, 76128 Karlsruhe, Germany}
  \\
  {\small\it (b) Theoretical Physics Department, CERN,}\\
  {\small\it 1211 Geneva, Switzerland}
  \\
  {\small\it (c) Physik-Institut, Universit\"at Z\"urich, }\\
  {\small\it Winterthurerstrasse 190, 8057 Z\"urich, Switzerland}
}

\date{}

\maketitle

\begin{abstract}

\noindent
We compute next-to-next-to-leading order QCD corrections to the  partonic
processes $b\to c \bar{u} d$ and $b\to c \bar{c} s$, which constitute 
the dominant decay channels in standard model predictions for $B$-meson lifetimes
within the heavy quark expansion.
We consider the contribution from the four-quark operators $O_1$ and $O_2$ in the $\Delta B =1$ effective Hamiltonian. 
The decay rates are obtained from the imaginary parts of four-loop 
propagator-type diagrams. 
We compute the corresponding master integrals using the ``expand and match'' approach which
provides semi-analytic results for the physical charm and bottom quark
masses. We show that the dependence of the decay rate on the renormalization
scale is significantly reduced after including the next-to-next-to-leading
order corrections.
Furthermore, we compute next-to-next-to-leading order corrections to the
Cabibbo-Kobayashi-Maskawa-suppressed decay channels $b \to u \bar c s$ and $b \to u \bar u d$.

\vspace*{5em}

\end{abstract}

\thispagestyle{empty}


\thispagestyle{empty}

\newpage


\section{Introduction} 

Bound states of quarks and anti-quarks in the form of hadrons are 
commonly observed in collider experiments. Their lifetimes are among the 
most important quantities to obtain
insights into the fundamental 
interactions of elementary particles. In this paper
we consider $B$ mesons which contain a heavy $b$ quark or anti-quark
and a lighter quark $u,d$ or $s$.  Their decay is governed by the weak
interaction of the $b$ quark. The theoretical framework describing the
decay rates of inclusive decays of hadrons containing a heavy quark is the
heavy quark expansion (HQE). Lifetimes, which are the inverse of the total decay width,
can be calculated in the HQE as a double series expansion in $\Lambda_{\mathrm{QCD}}/m_b$ and 
the strong coupling constant $\alpha_s$. 
The first term in the $\Lambda_{\mathrm{QCD}}/m_b$ expansion describes
the decay of a free bottom quark within perturbative QCD (for reviews see~\cite{Lenz:2014jha,Albrecht:2024oyn}).
The leading term is complemented by power-suppressed contributions involving 
matrix elements of higher-dimensional operators.
Global fits~\cite{Bernlochner:2022ucr,Finauri:2023kte} of $B \to X_c \ell \bar \nu_\ell$ 
data measured at $B$ factories
provide the numerical values for the matrix elements involving two-quark operators, 
like the kinetic and chromo-magnetic terms $\mu_\pi^2$ and $\mu_G^2$.
Lifetimes depend also on matrix elements of four-quark operators, 
which can be estimated by HQET sum rules~\cite{Kirk:2017juj,King:2021jsq} or 
calculated on the lattice~\cite{Lin:2022fun,Black:2023vju}.

At leading order in $\Lambda_{\mathrm{QCD}}/m_b$, 
the total decay width $\Gamma(B_q)$ is given by the sum of the semileptonic 
decay channels
$b \to c \ell \bar \nu_\ell$, with $\ell = e,\mu, \tau$, 
the nonleptonic channels $b\to c\bar{u}d$, $b\to c\bar{u}s$,
$b\to c\bar{c}d$ and $b\to c\bar{c}s$, as well as other CKM suppressed and rare decay modes.
For reviews we refer to Refs.~\cite{Lenz:2014jha,Albrecht:2024oyn}.
Next-to-next-to-leading order (NNLO) corrections to the semi-leptonic
decay rate have been computed a few years ago~\cite{Pak:2008qt,Pak:2008cp,Dowling:2008mc,Melnikov:2008qs}. 
More recently also the $O(\alpha_s^3)$ corrections~\cite{Fael:2020tow,Fael:2023tcv} became available, 
even for the kinematic moments without experimental cuts~\cite{Fael:2022frj}.

The nonleptonic decays of $B$ mesons are 
most conveniently
described with the help of the $\Delta B = 1$ effective Hamiltonian~\cite{Buras:1989xd,Buchalla:1995vs,Buras:2011we}
governing the low-energy dynamics at the renormalization scale $\mu_b \sim m_b$.
For them only next-to-leading order
(NLO) corrections are currently available~\cite{Altarelli:1991dx,Buchalla:1992gc,Bagan:1994zd,Bagan:1995yf,Greub:2000sy,Greub:2000an,Krinner:2013cja}.
At NNLO there are only partial results from Ref.~\cite{Czarnecki:2005vr}
where only one of the relevant four-quark operators has been
considered. Furthermore, no resummation of the large logarithms $\log(M_W/m_b)$
due to the running from the electroweak scale to the scale of the $B$ meson mass has been performed.

On the basis of all currently available correction terms,
one obtains the following results for the $B$ meson decay rates~\cite{Lenz:2022rbq}
\begin{eqnarray}
  \Gamma(B^+) &=& 0.58^{+0.11}_{-0.07} \, \mbox{ps}^{-1}\,,\nonumber\\[10pt]
  \Gamma(B_d) &=& 0.63^{+0.11}_{-0.07} \, \mbox{ps}^{-1}\,,
  \label{eq::Bdec}
\end{eqnarray}
with an uncertainly of almost 20\%. It is by far dominated by the
renormalization scale dependence of the free-quark decay. 
The uncertainties arising from CKM elements and 
quark mass values are significantly smaller. 
For this reason, the current state-of-the-art calls for a determination of NNLO corrections
to nonleptonic decays in the free quark approximation, 
including an appropriate choice of the short-distance mass scheme for the heavy quarks. 
In this work, we aim to address this gap by providing results for 
the NNLO corrections to the nonleptonic decay of a free quark, 
where the charm and bottom quark masses are renormalized on-shell.
We take into account finite charm and bottom quark masses
and consider the so-called current-current operators $O_1$ and $O_2$,
which provide the dominant contribution to the decay width. We
will show that the $\mu_b$ dependence is significantly reduced once the
NNLO corrections are included.  For an update of the decay widths in
Eq.~(\ref{eq::Bdec}) it is necessary to consider other renormalization
schemes for the quark masses. Furthermore, one has to properly combine all
decay channels and incorporate the known power-suppressed terms~\cite{Beneke:2002rj,Franco:2002fc,Gabbiani:2004tp,Gabbiani:2003pq,Lenz:2020oce,Mannel:2023zei}. 
This is postponed to a future publication~\cite{in_prep}.

The paper is organised as follows: 
In Section~\ref{sec::framework} we set up the notation, introduce the 
effective Hamiltonian and the operators $O_1$ and $O_2$. 
We discuss in particular how to apply naive dimensional regularization
and use anticommuting $\gamma_5$. This is crucial in order to adopt the same 
prescription utilized in the calculation of the NNLO anomalous dimensions
of $O_1$ and $O_2$~\cite{Buras:1989xd,Gorbahn:2004my}.
The detail of the calculation of interference terms up to $O(\alpha_s^2)$
and the evaluation of the four-loop master integrals are presented in
Section~\ref{sec:technicaldetails}. 
We also discuss in details the role of evanescent operators in the calculation.
In Section~\ref{sec:results} we combine our predictions for the squared amplitudes
up to $O(\alpha_s^2)$ with the NNLO Wilson coefficients evaluated at 
the low-scale $\mu_b \sim m_b$ and give results for the rate of the different
channels. We conclude in Section~\ref{sec:conclusions}.
In the Appendix~\ref{sec::OpMix}, we provide additional details about the 
operator renormalization.


\section{\label{sec::framework}Framework}

We describe the nonleptonic decays of a bottom quark governed by weak interactions
using the effective Hamiltonian
\begin{equation}
  \mathcal{H}_{\rm eff} =
  \frac{4 G_F}{\sqrt{2}}
  \sum_{q_{1,3} = u,c}
  \sum_{q_2 = d,s}
  \lambda_{q_1 q_2 q_2}
  \Big(
    C_1(\mu_b) O_1^{q_1 q_2 q_3}
    +C_2(\mu_b) O_2^{q_1 q_2 q_3}
  \Big)
  +{\rm h.c.} \, ,
  \label{eqn:Heff}
\end{equation}
where $\lambda_{q_1 q_2 q_3} = V_{q_1 b} V_{q_2 q_3}^\star$ are the
corresponding CKM matrix elements and $C_i(\mu_b)$ are the
Wilson coefficients for the $\Delta B = 1$ effective operators evaluated at
the renormalization scale $\mu_b \sim m_b$.  The current-current operators
$O_i^{q_1 q_2 q_3}$ are given by~\cite{Buras:1989xd,Buchalla:1995vs}
\begin{align}
  O_1^{q_1 q_2 q_3} &=
  (\bar q_1^\alpha \gamma^\mu P_L b^\beta) (\bar q_2^\beta \gamma_\mu P_L q_3^\alpha),
  \notag \\
  O_2^{q_1 q_2 q_3} &=
  (\bar q_1^\alpha \gamma^\mu P_L b^\alpha) (\bar q_2^\beta \gamma_\mu P_L q_3^\beta)\,,
  \label{eqn:O1O2definition}
\end{align}
where $\alpha$ and $\beta$ refer to colour indices.  
We will refer to such operator definition as the \textit{historical} basis.  For simplicity, we will ignore
the penguin operators whose contributions to the rate are suppressed
due to the numerically small Wilson coefficients.  Another common
operator choice is the so-called \textit{Chetyrkin-Misiak-M\"unz} (CMM) basis~\cite{Chetyrkin:1997gb}, in which
the operators are
\begin{align}
  O_1^{' \, q_1 q_2 q_3} &=
  (\bar q_1 T^a \gamma^\mu P_L b) (\bar q_2 T^a \gamma_\mu P_L q_3),
  \notag \\
  O_2^{' \, q_1 q_2 q_3} &=
  (\bar q_1 \gamma^\mu P_L b) (\bar q_2 \gamma_\mu P_L q_3),
  \label{eqn:O1O2CMMdefinition}
\end{align}
where $T^a$ are the generator of the $SU(3)$ colour group.
The CMM basis was introduced to consistently use fully anticommuting $\gamma_5$ at any number of loops
in the evaluation of the QCD corrections to $b \to s \gamma$ and $b \to s \ell \ell$ decays.
However, this feature breaks down for the processes considered in this article and
the historical basis turns out to be more convenient for our calculation, as explained below.
Moreover, the historical basis is the default choice in many phenomenological studies~\cite{Lenz:2014jha,Lenz:2020oce,Lenz:2022rbq,Lenz:2022pgw,Albrecht:2024oyn}.

\begin{figure}[t]
  \centering
  \begin{subfigure}{0.3\textwidth}
  \includegraphics[width=\textwidth]{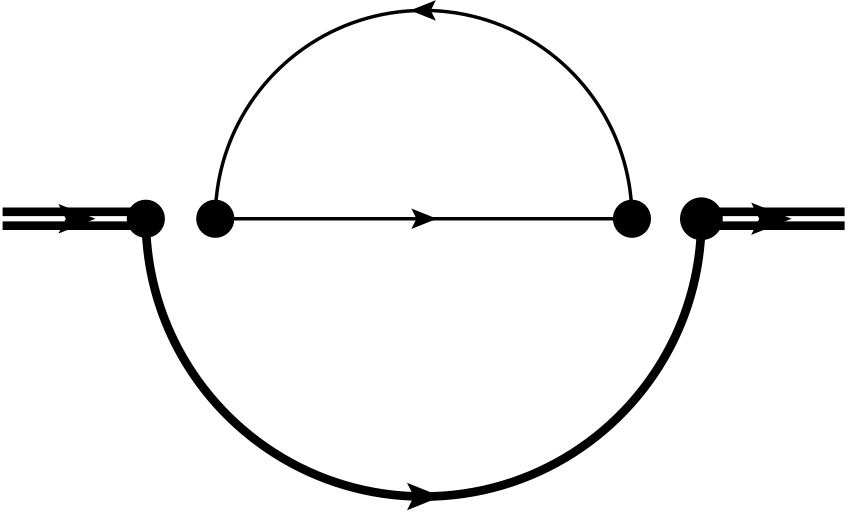}
  \caption{}
  \end{subfigure}
  \begin{subfigure}{0.3\textwidth}
  \includegraphics[width=\textwidth]{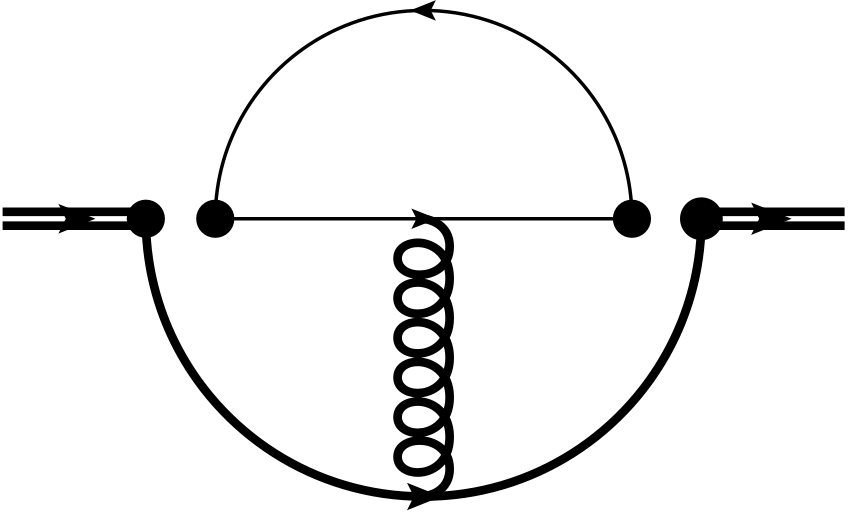}
  \caption{}
  \end{subfigure}
  \begin{subfigure}{0.3\textwidth}
  \includegraphics[width=\textwidth]{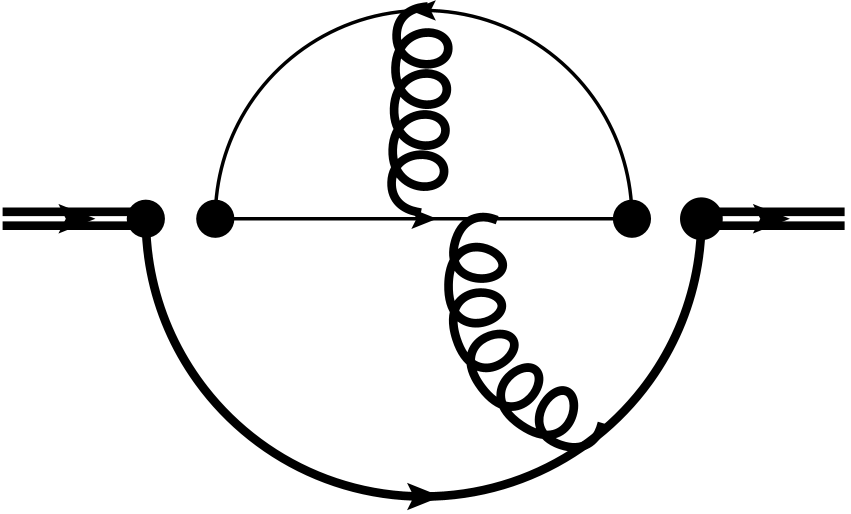}
  \caption{}
  \end{subfigure}
  \begin{subfigure}{0.3\textwidth}
  \includegraphics[width=\textwidth]{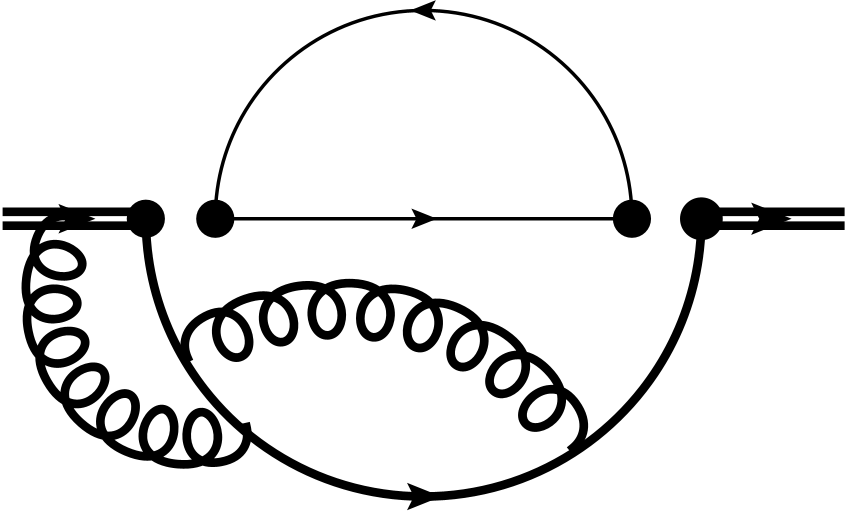}
  \caption{}
  \end{subfigure}
  \begin{subfigure}{0.3\textwidth}
  \includegraphics[width=\textwidth]{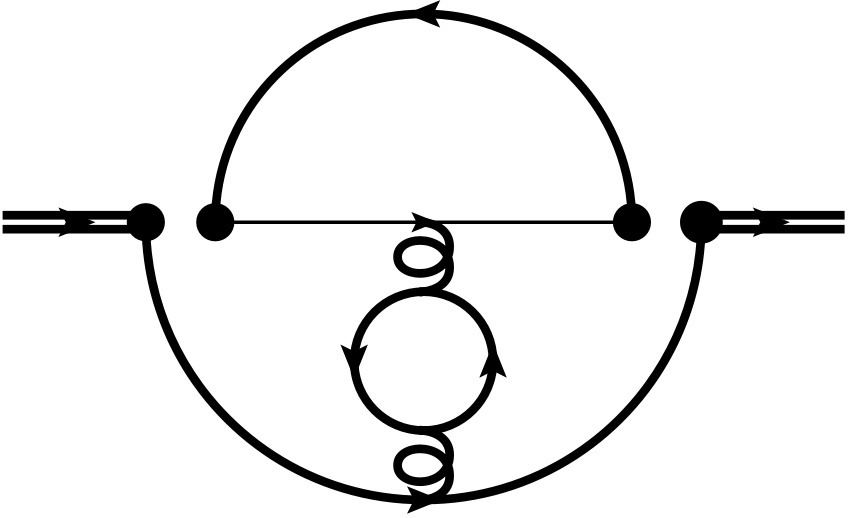}
  \caption{}
  \end{subfigure}
  \begin{subfigure}{0.3\textwidth}
  \includegraphics[width=\textwidth]{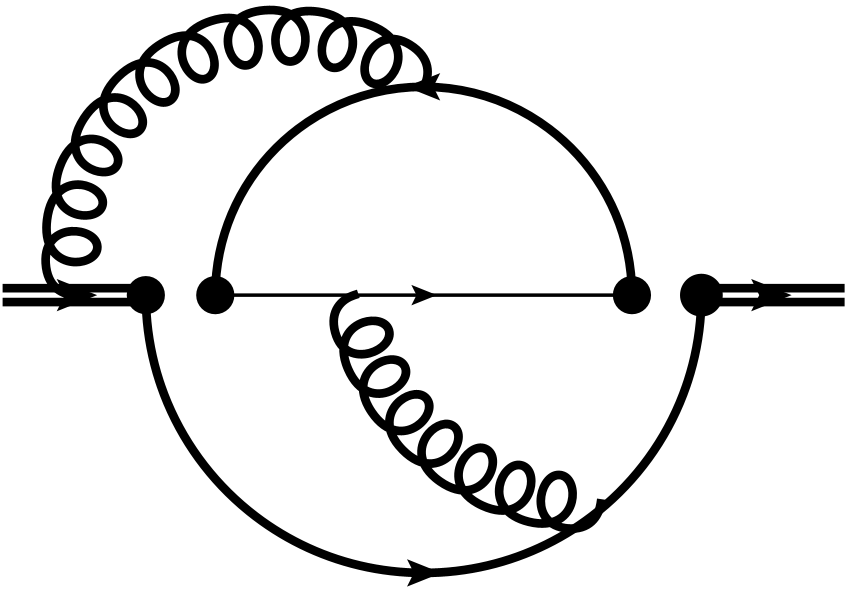}
  \caption{}
  \end{subfigure}
  \begin{subfigure}{0.3\textwidth}
  \includegraphics[width=\textwidth]{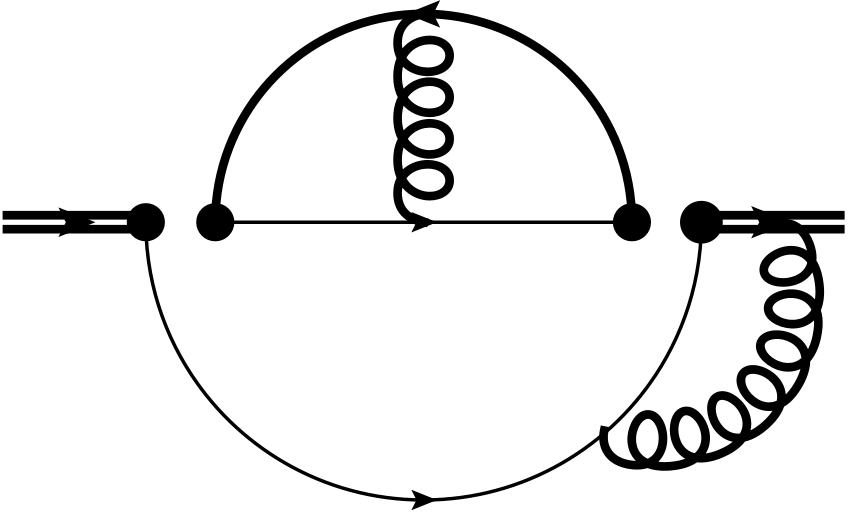}
  \caption{}
  \end{subfigure}
  \begin{subfigure}{0.3\textwidth}
  \includegraphics[width=\textwidth]{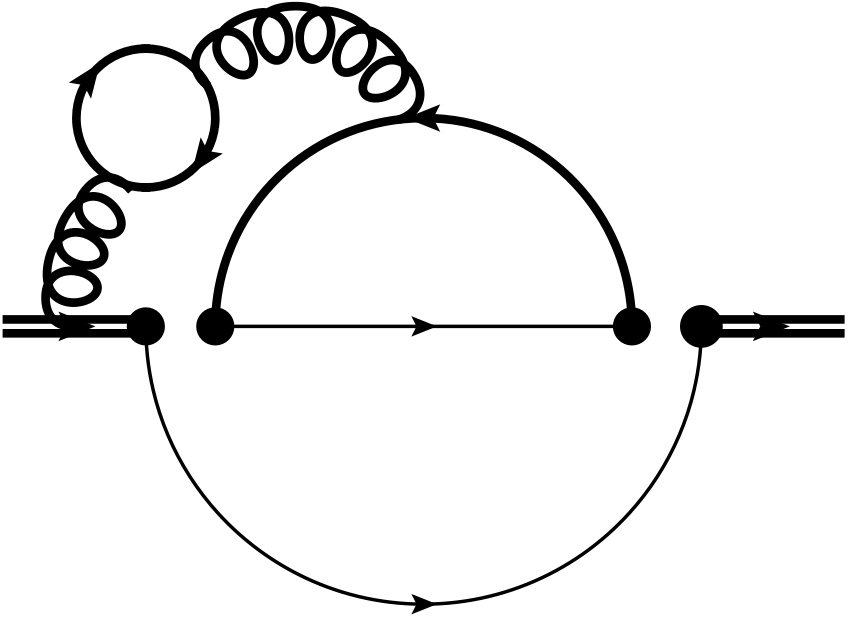}
  \caption{}
  \end{subfigure}
  \begin{subfigure}{0.3\textwidth}
  \includegraphics[width=\textwidth]{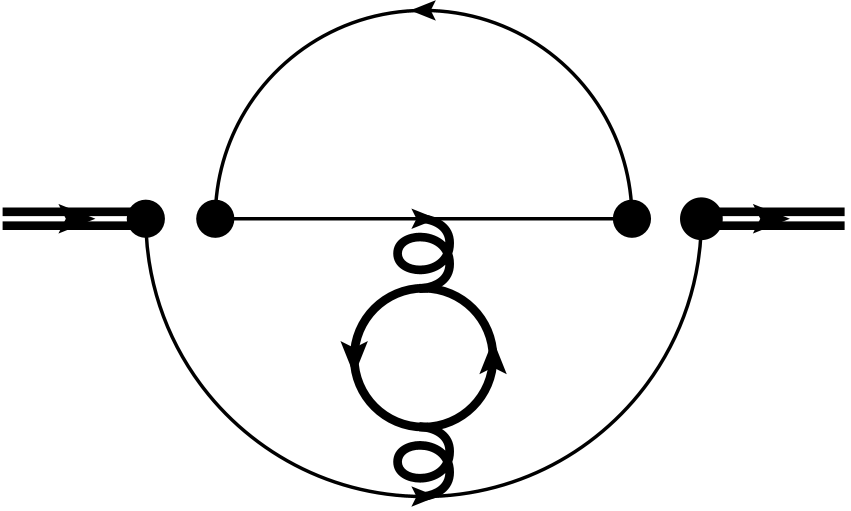}
  \caption{}\label{fig:Uc}
  \end{subfigure}
  \caption{Example of Feynman diagrams contributing to the forward scattering amplitude
  in Eq.~\eqref{eqn:Gamma} at order  $\alpha_s^0$ (LO), $\alpha_s$ (NLO) and $\alpha_s^2$ (NNLO).
  The dot pairs denote insertions of the effective operators and
highlight how fermion flows are contracted. Thin, bold and double bold lines denote massless, charm and bottom propagators.}
  \label{fig:NLOnnlep}
\end{figure}

In our study, we treat the bottom and the charm quark as massive with mass $m_b$ and $m_c$, respectively,
while all other quarks are considered massless ($m_{u,d,s} = 0$).
We then divide the nonleptonic decays into three classes
based on the flavour indices of $O_1$ and $O_2$:
\begin{enumerate}
    \item[(i):] Three massless quarks in the final state, i.e.\  $q_1 q_2 q_3 = u d u, u s u$.
    \item[(ii):] One charm quark and two massless quarks ($q_1 q_2 q_3 = c d u, c s u, u d c, u s c$).
    \item[(iii):] Two charm quarks and one massless quark ($q_1 q_2 q_3 = c d c$ and $c s c$).
\end{enumerate}
In the following, we will focus on the CKM-favoured decays $b\to c \bar u d$ 
and the CKM-suppressed mode $b \to u \bar c s$ as representatives for case (ii).
For case (iii) we consider $b \to c \bar c s$.
Case (i) can be obtained from the $m_c \to 0$ limit of the other two,
however it requires the additional calculation of the finite charm-mass effect originating
from closed charm-loop insertion into a gluon propagator~(see the sample diagram in Fig.~\ref{fig:NLOnnlep}(i)).
We will refer to this kind of effect as the $U_c$ contribution.
Note also that our NNLO results will include (small) contributions associated to the production 
of an addition $\bar c c$ pair from gluon splitting, e.g.\ $b \to c\bar{u}d (g^\star \to \bar cc)$.

\begin{figure}[t]
  \centering
  \begin{subfigure}{0.3\textwidth}
  \includegraphics[width=\textwidth]{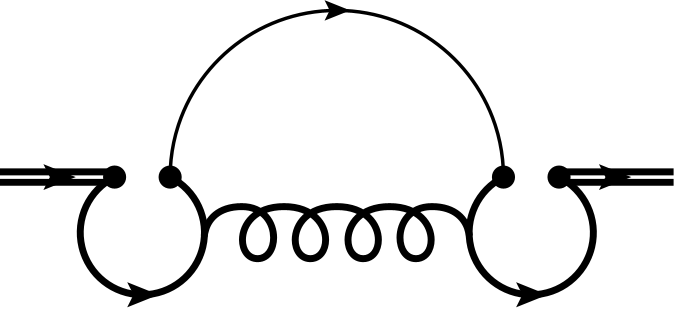}
  \caption{}
  \end{subfigure}
  \begin{subfigure}{0.3\textwidth}
  \includegraphics[width=\textwidth]{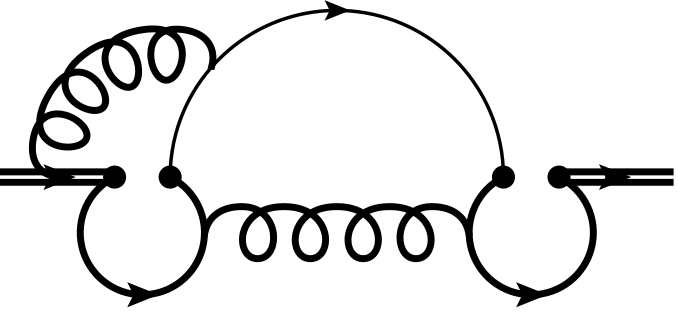}
  \caption{}
  \end{subfigure}
  \begin{subfigure}{0.35\textwidth}
  \includegraphics[width=\textwidth]{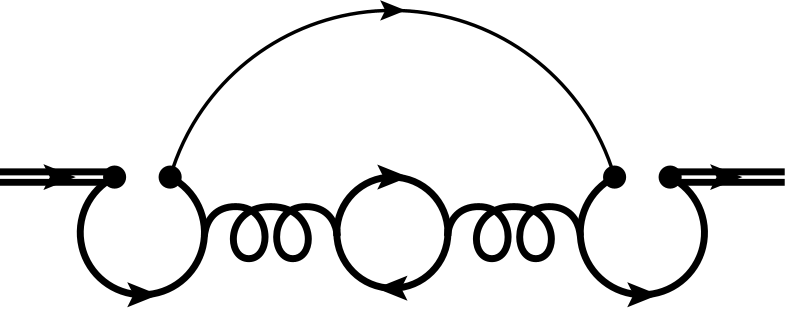}
  \caption{}
  \end{subfigure}
  \caption{Example of Feynman diagrams with penguin topology at NLO and NNLO.}
  \label{fig:NLOnnlepPenguin}
\end{figure}

To calculate the inclusive decay width, we use the optical theorem and
evaluate the imaginary part of forward scattering amplitudes for an
on-shell bottom quark up to NNLO.  For a definite flavour content of the
operators, the contribution to the decay rate can be written as
\begin{multline}
  \Gamma^{q_1q_2q_3}(\rho) =
  \frac{1}{m_b} 
  \sum_{i,j = 1,2}
  \left(\frac{4 G_F |\lambda_{q_1q_2q_3}|}{\sqrt{2}} \right)^2
  \\
  C_i^\dagger(\mu_b) C_j(\mu_b) \,
  {\rm Im} \, 
  i \int {\rm d}^4x \,e^{{\rm i}qx}\,
  \bra{b} 
  T\Big\{ O_i^{\dagger \, q_1q_2q_3 }(x)  O_j^{q_1q_2q_3}(0) \Big\}
  \ket{b} \Bigg|_{q^2=m_b^2}\,,
  \label{eqn:Gamma}
\end{multline}
where $\rho = m_c/m_b$.
As a consequence, at LO the imaginary parts of two-loop diagrams have to be
computed and at NLO and NNLO, three- and four-loop diagrams have to be
considered, see Fig.~\ref{fig:NLOnnlep}.
For $b \to c \bar c s$ and $b \to u \bar u d$, besides the corrections
in which the LO diagram in Fig.~\ref{fig:NLOnnlep}(a) is dressed
with additional gluon lines, there are also contributions
at order $O(\alpha_s)$ and $O(\alpha_s^2)$ due to the insertions
of the operators $O_{1,2}$ into penguin diagrams like Fig.~\ref{fig:NLOnnlepPenguin}.
These kind of corrections of $O(\alpha_s)$ were studied
in~\cite{Lenz:1998qp,Lenz:1997aa,Krinner:2013cja} and shown to be numerically much smaller
than the $O(\alpha_s)$ corrections arising from diagrams like
Fig.~\ref{fig:NLOnnlep}(b).
We postpone the evaluation of this class of penguin-like
diagrams to a subsequent publication since they require a
special treatment of cut Feynman integrals.

Due to the presence of $\gamma_5$ a straightforward use of naive dimensional
regularization (NDR) is not possible. Starting from NLO there are traces with one
$\gamma_5$ matrix that must be evaluated in $d=4-2\epsilon$ dimensions (see
e.g.\ the diagram in Fig.~\ref{fig:NLOnnlep}(b)). For the calculation of the
anomalous dimension (and thus the renormalization constants needed for the
operator mixing) anticommuting $\gamma_5$ has been
used~\cite{Buras:1989xd,Gorbahn:2004my}. Thus we would like to apply the same prescription
in our calculation.

\begin{figure}[t]
  \centering
  \includegraphics[width=\textwidth]{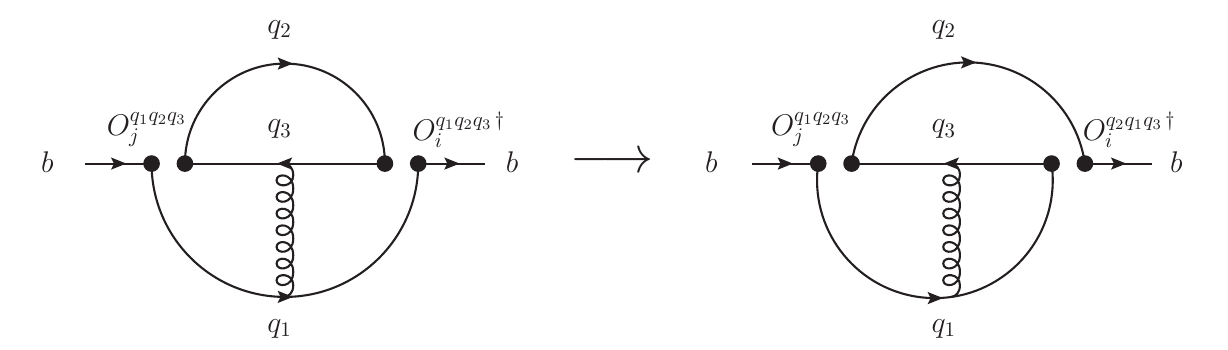}
  \caption{Illustration of the Fierz transformation on a diagram at NLO.
  In contrast to the diagram on the left contributing to $\Gamma^{q_1q_2q_3}$
  in Eq.~\eqref{eqn:Gamma}, the one on the right can be computed using
  anticommuting $\gamma_5$.
  }
  \label{fig:Fierz}
\end{figure}

In order to use anticommuting $\gamma_5$, we apply a method similar to the
one discussed in Sec.~2.3 of Ref.~\cite{Bagan:1994zd}. Instead of
Eq.~\eqref{eqn:Gamma}, let us consider
\begin{multline}
  \widetilde{\Gamma}^{q_1q_2q_3}(\rho) 
  =
  \frac{1}{m_b} 
  \sum_{i,j = 1,2}
  \left(\frac{4 G_F |\lambda_{q_1q_2q_3}|}{\sqrt{2}} \right)^2
  \\
  \widetilde{C}_i^\dagger(\mu_b) C_j(\mu_b) \,
  {\rm Im} \, 
  i \int {\rm d}^4x \, e^{{\rm i}qx}
  \bra{b} 
  T\Big\{ O_i^{q_2 q_1 q_3 \, \dagger}(x) O_j^{q_1q_2q_3}(0) \Big\}
  \ket{b}
  \Bigg|_{q^2=m_b^2}\,.
  \label{eqn:Gammatilde}
\end{multline}
Note the different ordering of the quark flavour indices in the first operator, i.e.\ $q_2 q_1 q_3$ instead of 
$q_1 q_2 q_3$.  
Here, $\widetilde{C}_i(\mu_b)$ is the Wilson coefficient of the operator
$O_i^{q_2 q_1 q_3 }$.  
The forward scattering matrix defined by $\widetilde{\Gamma}^{q_1q_2q_3}$
leads to only one trace of gamma matrices containing an odd number of $\gamma_5$. 
This is easily seen from Fig.~\ref{fig:Fierz} where the effect is illustrated. 
Since the width is parity-even, the trace containing exactly one $\gamma_5$ can be discarded at any order in perturbation theory, 
while in case we encounter two $\gamma_5$ in the same trace we apply anticommuting $\gamma_5$.
This means that for $\widetilde{\Gamma}^{q_1q_2q_3}$ we can use NDR and the renormalization constants known from the literature.
If we had considered $\Gamma^{q_1q_2q_3}$, we would have encountered the product of two traces 
with one $\gamma_5$ each, which in general can give a parity-even contribution. 
This is shown in the NLO diagram on the left in Fig.~\ref{fig:Fierz}. 

In the next step we recover the expression for $\Gamma^{q_1 q_2 q_3}$ 
from $\widetilde \Gamma^{q_2 q_1 q_3}$.  In four dimensions
the operators $O_1^{q_2q_1q_3}$ and $O_2^{q_1q_2q_3}$ are connected
by a Fierz transformation\footnote{We include also the factor $-1$ for anti-commuting fields.}:
\begin{align}
  O_1^{q_1q_2q_3} &=
  (\bar q_1^\alpha \gamma^\mu P_L b^\beta) (\bar q_2^\beta \gamma_\mu P_L q_3^\alpha)
  = (\bar q_2^\alpha \gamma^\mu P_L b^\alpha) (\bar q_1^\beta \gamma_\mu P_L q_3^\beta)
  = O_2^{q_2q_1q_3},
\end{align}
and similarly $O_2^{q_1q_2q_3} = O_1^{q_2q_1q_3}$. This implies
\begin{equation}
  \Gamma^{q_1 q_2 q_3} (\rho)=
  \widetilde \Gamma^{q_2 q_1 q_3} (\rho)
  \Big\vert_{\widetilde C_1 \to C_2, \, \widetilde C_2 \to C_1}.
  \label{eqn:Fierzrelation}
\end{equation}
Fierz identities in general are not valid for $d \neq 4$, 
however, the Fierz symmetry can be restored order by order in perturbation theory by renormalization
using anticommuting $\gamma_5$ and a suitable definition of the evanescent operators~\cite{Buras:1989xd}.
We will discuss in detail the renormalization and the evanescent operator scheme in section~\ref{sec:renormalization}.
In conclusion, our strategy for obtaining the NNLO prediction for $\Gamma^{q_1 q_2 q_3}$ is to calculate
$\widetilde \Gamma^{q_1 q_2 q_3}$ and to adopt a renormalization scheme which preserves Eq.~\eqref{eqn:Fierzrelation} 
up to $O(\alpha_s^2)$.


\section{Technical details}
\label{sec:technicaldetails}
In this Section we provide details for the calculation of the squared amplitude, the master integrals and the 
renormalization scheme adopted for the effective operators in Eq.~\eqref{eqn:O1O2definition}.
\subsection{Generation of the amplitude}

For our calculation we use a well-tested chain of programs
which allows for a high degree of automation.  We us {\tt
  qgraf}~\cite{Nogueira:1991ex} for the generation of the amplitude and {\tt
  tapir}~\cite{Gerlach:2022qnc} for the translation to {\tt FORM}~\cite{Kuipers:2012rf} code and
the identification of the underlying integral families. The program {\tt exp}~\cite{Harlander:1998cmq,Seidensticker:1999bb}
performs the mapping of the amplitudes to the integral families and prepares
them for further processing with {\tt FORM}.
After applying projectors and decomposing the reducible numerator factors in
terms of denominators, we obtain for each family a list of scalar integrals
for which we need a reduction to so-called master integrals.
For the decay channels with up to one charm quark in the final state at LO we
perform the calculation for general QCD gauge parameter
keeping linear terms and
check that it drops out at the level of the renormalized amplitude.
For the channel $b\to c\bar{c}s$ we choose Feynman gauge since the reduction
is significantly more expensive.

We employ the program {\tt Kira}~\cite{Klappert:2020nbg} in combination with {\tt Fermat}~\cite{fermat} and \texttt{FireFly}~\cite{Klappert:2019emp,Klappert:2020aqs}.
for the reduction to master integrals, which is
organised in two steps. First, we generate (for each family) reduction tables
for seed integrals with up to two dots and one scalar product up to the top-level sector.
These reduction tables serve as input for the program {\tt
  ImproveMasters.m}~\cite{Smirnov:2020quc} to search for a \textit{good basis}, i.e.\ 
a master integral basis where the dependence on the kinematical quantity $\rho$ and on the dimension $d$ factorizes 
in all denominators.  
In the second step, we perform the reduction of the integrals needed for
the amplitude onto the good basis that we found. We use {\tt Kira} also for the minimization of the master
integrals among all families.
For the process $b\to c \bar{u}d$ we find 321, for
$b\to c \bar{c}s$ 527 and for $U_C$ 21 master integrals.
The calculation of the amplitudes for $b\to u\bar{c}s$ can be mapped to the same families as $b\to c\bar{u}d$.


\subsection{Computation of the master integrals}

\begin{figure}[t]
  \centering
  \begin{subfigure}{0.3\textwidth}
  \includegraphics[width=\textwidth,valign=c]{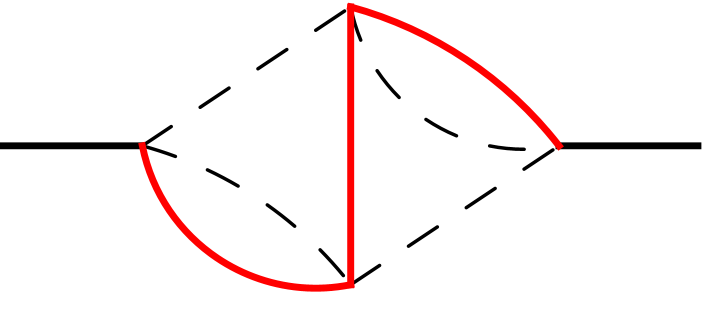}
  \end{subfigure}
  \begin{subfigure}{0.3\textwidth}
  \includegraphics[width=\textwidth,valign=c]{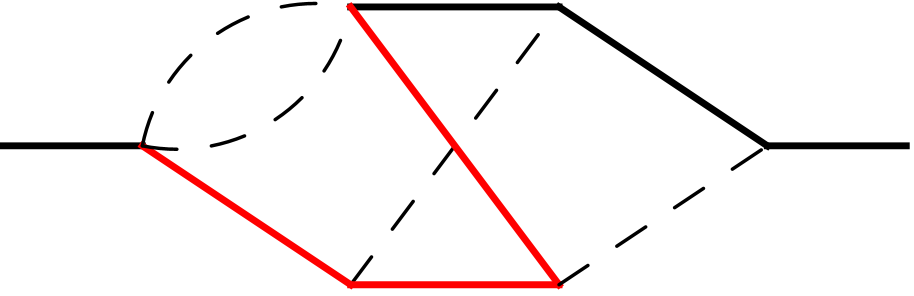}
  \end{subfigure}
  \begin{subfigure}{0.3\textwidth}
  \includegraphics[width=\textwidth,valign=c]{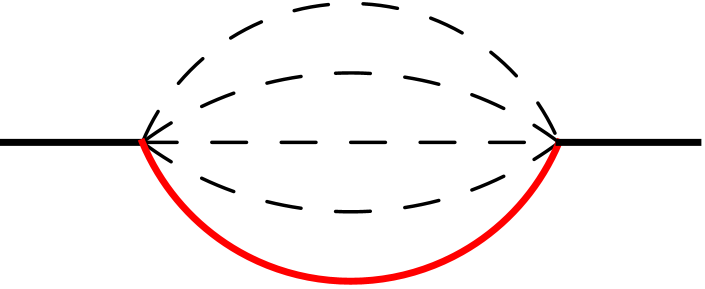}
  \end{subfigure}\\[10pt]
  \begin{subfigure}{0.3\textwidth}
  \includegraphics[width=\textwidth,valign=c]{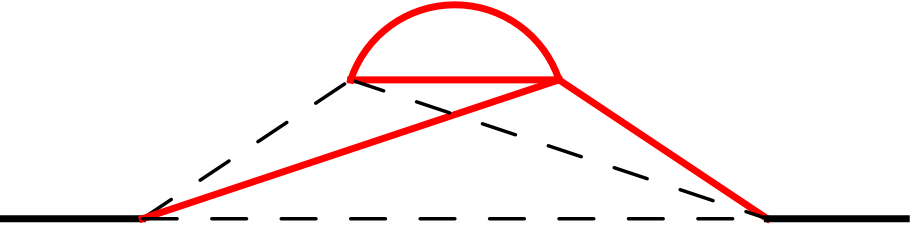}
  \end{subfigure}
  \begin{subfigure}{0.3\textwidth}
  \includegraphics[width=\textwidth,valign=c]{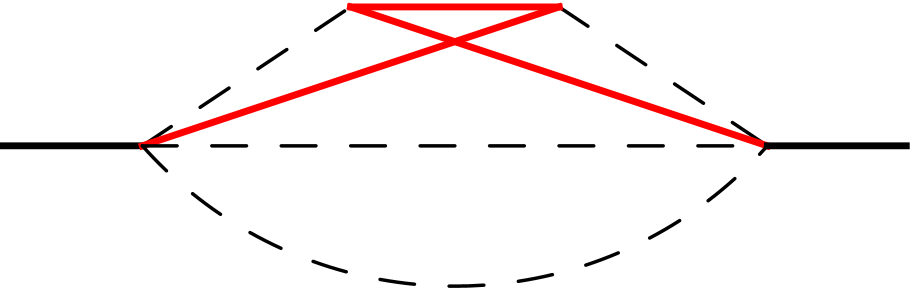}
  \end{subfigure}
  \begin{subfigure}{0.3\textwidth}
  \includegraphics[width=\textwidth,valign=c]{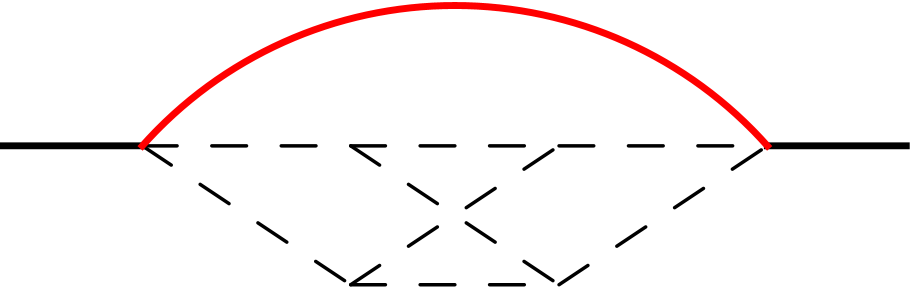}
  \end{subfigure}\\[10pt]
  \begin{subfigure}{0.3\textwidth}
  \includegraphics[width=\textwidth,valign=c]{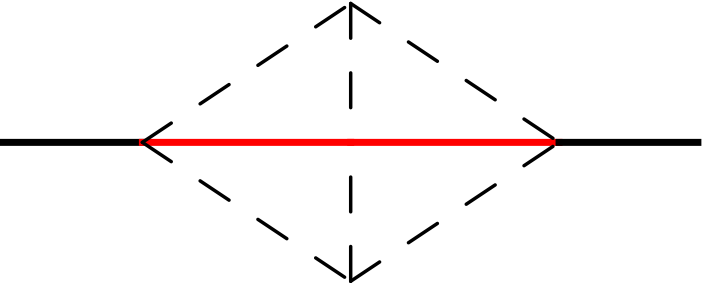}
  \end{subfigure}
  \begin{subfigure}{0.3\textwidth}
  \includegraphics[width=\textwidth,valign=c]{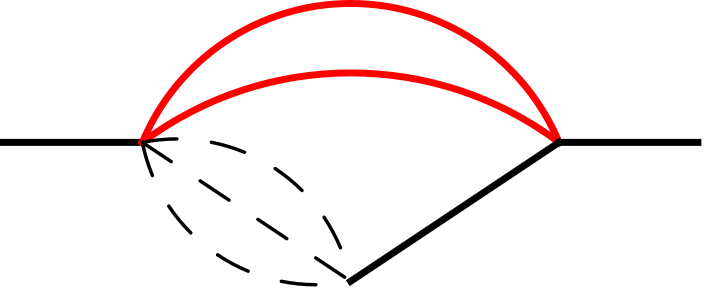}
  \end{subfigure}
  \begin{subfigure}{0.3\textwidth}
  \includegraphics[width=\textwidth,valign=c]{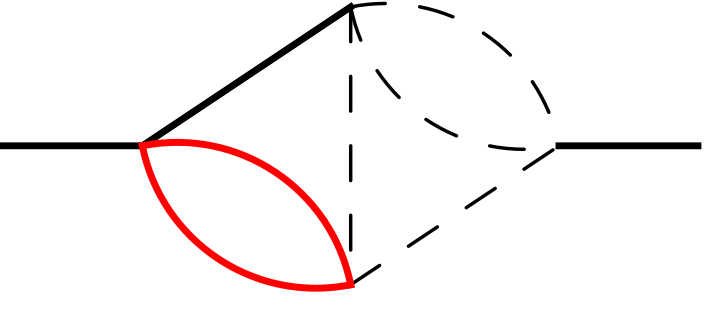}
  \end{subfigure}
  \caption{Samples of four-loop master integrals. Black and red solid lines represents
massive propagator with mass $m_b$ and $m_c$, respectively, while dashed lines are massless
propagators.}
  \label{fig:MIsnnlep}
\end{figure}

The master integrals at LO and NLO are calculated analytically. 
For both channels  
$b\to c \bar u d$ and $b\to c \bar c s$,
we establish a set of differential equations for the master integrals
by differentiating them with respect to $\rho$. 
Afterwards, we use the programs {\tt Canonica}~\cite{Meyer:2017joq} and \texttt{Libra}~\cite{Lee:2020zfb} (the latter implements Lee’s algorithm~\cite{Lee:2014ioa}) to find a suitable basis transformation 
such that the masters in the new basis satisfy a set of differential equations in canonical form~\cite{Henn:2013pwa}.
The boundary conditions to the differential equations are obtained using the auxiliary mass flow method \cite{Liu:2017jxz,Liu:2021wks} as implemented in \texttt{AMFlow} \cite{Liu:2022chg}.
We numerically compute all master integrals at NLO at a regular value of $\rho$
with about 150 digits, a precision sufficient to reconstruct 
the boundary constants in terms of transcendental numbers using the \texttt{PSLQ} algorithm~\cite{pslq}.
The master integrals for $b\to c \bar u d$ are expressed through simple
Harmonic Polylogarithms (HPLs)~\cite{Remiddi:1999ew} with $\rho$ as argument.
For the decay $b\to c \bar c s$, we apply the change of variable
\begin{equation}
    \rho = \frac{x}{1+x^2} ,
\end{equation}
to bring the system in canonical 
form. The solution is expressed in terms of 
iterated integrals over the letters $x,1+x,1-x,1+x^2,1 - x + x^2$ and $1 +x + x^2$, which are also known as cyclotomic harmonic polylogarithms~\cite{Ablinger:2011te}.
After factorizing the letters over the complex numbers, we can express the results in terms of
generalized polylogarithms (GPLs)~\cite{Goncharov:1998kja,Goncharov:2001iea}.

At four-loop order we exclusively use the ``expand and match''
method~\cite{Fael:2021kyg,Fael:2022rgm,Fael:2022miw,Fael:2023zqr} which
provides semi-analytic results for the master integrals in the form of expansions around 
properly chosen points with numerical coefficients. 
We use ``expand and match'' at LO and NLO as cross check for the analytic result. 
At NNLO the application 
to the subset of master integrals which already contribute 
the semileptonic decay rate is described in detail 
in Ref.~\cite{Egner:2023kxw}.
In the following we concentrate on a brief summary and 
on the additional features
present in the hadronic decay rate.
In principle we could concentrate on the physical
region with $m_c/m_b\approx 0.3$. However, for $b\to c\bar{u}d$ and $U_C$ we
want to cover the whole region for $0\le m_c/m_b\le1$
so that our results can also be applied to
other physical situations.
In the case of $b\to c\bar{c}s$, which is computationally much more demanding, we construct expansions which provide precise results for $0\le m_c/m_b \lesssim 0.4$.

One of the ingredients for ``expand and match'' 
are the differential equations for the master integrals
in the variable
\begin{equation}
  \rho=\frac{m_c}{m_b}\,.
\end{equation}
Let us stress that the system of differential equations does not have to be in a particular form;
in particular it is not necessary to bring it into a canonical
form~\cite{Henn:2013pwa,Lee:2014ioa}. However, the computational
time can be reduced in case the occurring denominators have a simple
structure.

We denote the positions of the poles in the differential equation
as singular points. In general, such branch cut points in the complex plane 
reflect physical thresholds. Some of the singularities are also spurious
and no divergent behaviour is observed in the amplitude.
All other points are called regular since there the master integrals
have usual Taylor expansions.

As further ingredient, we need boundary conditions in the form of 
analytic or high-precision numerical results for the master integrals
at a regular point of the differential equation.
In our application we obtain the numerical boundary conditions
with the help of \texttt{AMFlow}~\cite{Liu:2022chg}, typically requesting
80~digits. For $b\to c\bar{u}d$ we evaluate them for $\rho=1/4$,
for $b\to c\bar{c}s$ at $\rho=1/5$ and $\rho=1/3$, and 
for $U_C$ at $\rho=1/3$.

The basic idea of ``expand and match'' is to construct truncated expansions around
regular or singular points with the help of the differential equations and
match them numerically at intermediate $\rho$-values.  The first expansion point
coincide with the value where the boundary conditions are computed.

At four-loop order we have singular points for those
values of $\rho$ corresponding to the thresholds for the production of one, two, three or four charm quarks. 
For the various channels we have:
\begin{align}
   \rho_{\mathrm{singular}} & \in  \{0,1/3,1\}   
   &  \text{ for } & b\to c\bar{u}d\,,\nonumber\\
   \rho_{\mathrm{singular}}& \in  \{0,1/4,1/2\} 
   &  \text{ for } & b\to c\bar{c}s\,,\nonumber\\
   \rho_{\mathrm{singular}}& \in  \{0,1/2\}     
   &  \text{ for } & U_C\,.
\end{align}

In the ``expand and match'' approach, usually the expansions around
singular points are necessary in order to connect the
solutions above and below the singularity. In order
to improve the precision one can complement the
list of expansion points by regular points. In our
case we choose
\begin{eqnarray}
    \rho_0\in\{0,1/4,1/3,1/2,7/10,1\} && \text{for } b\to c\bar{u}d\,,\nonumber\\
    \rho_0\in\{0, 1/5,1/3\} && \text{for } b\to c\bar{c}s\,,\nonumber\\
        \rho_0\in\{0,1/3,1/2,7/10,1\} && \text{for } U_C\,.
\end{eqnarray}
The combination of the expansions provided for $b\to c\bar{u}d$ and $U_C$ 
lead to a precise covering of the whole range for $\rho\in [0,1]$. In the case of $b\to c\bar{c}s$ we can cover the physically interesting region $\rho\in[0.2,0.4]$ with high precision. Furthermore, as we will see below, both Taylor expansions agree at the singular point $\rho_0=1/4$ to about 15 digits and thus the computation of the expansions around $\rho_0=1/4$, which is computationally quite expensive, can be avoided for practical purposes. Nonetheless, also for $b \to c \bar c s$ we provide for convenience an expansion around the massless charm quark limit.

Let us briefly discuss the different ans\"atze which we
have to use for the different expansion points.
They have to incorporate the respective physical situation
and contain logarithms and/or square roots.

For regular points $\rho_0$ the ansatz for the master integral $I_i$ is a simple Taylor expansion
and it is given by
\begin{eqnarray}
  I_i(\rho,\epsilon) &=&
  \sum_{j=-4}^{\epsilon_{\mathrm{max}}}
  \sum_{n=0}^{n_{\mathrm{max}}}
  c_{i,j,n} \epsilon^j \left(\rho - \rho_0 \right)^n
  \,.
  \label{eq::TaylorAnsatz}
\end{eqnarray}
For $\rho_0=0$ a power-log expansion is needed which we parametrize as 
\begin{eqnarray}
  I_i(\rho, \epsilon) &=&
  \sum_{j=-4}^{\epsilon_{\mathrm{max}}}
  \sum_{m=0}^{j+4}  \sum_{n=0}^{n_{\mathrm{max}}}
  c_{i,j,m,n} \epsilon^j \, (\rho-\rho_0)^n \, \log^m\left(\rho-\rho_0\right)\,.
  \label{eqLLAnsatzdel0}
\end{eqnarray}
We can use this ansatz also for threshold singularities involving
an odd number of particles. For an
even number of cut particles, the ansatz has to contain
also square roots~\cite{Davydychev:1999ic} and reads
\begin{eqnarray}
  I_i(\rho,\epsilon) &=&
  \sum_{j=-4}^{\epsilon_{\mathrm{max}}}\sum_{m=0}^{j+4}
  \sum_{n=n_{\mathrm{min}}}^{n_{\mathrm{max}}}
  c_{i,j,m,n} \epsilon^j \,  (\rho -\rho_0)^{n/2} \, 
   \log^m\left(\rho-\rho_0\right)\,.
\label{eq::powerlog}
\end{eqnarray}
We use this ansatz for $\rho_0=1/2$.


\subsection{Renormalization}
\label{sec:renormalization}

In the following we discuss 
the renormalization scheme adopted for the operators $O_1$ and $O_2$ 
to preserve the Fierz symmetry in Eq.~\eqref{eqn:Fierzrelation}
up to NNLO.

In a first step we perform the usual parameter renormalization for the strong
coupling constant in the $\overline{\rm MS}$ scheme with five active flavours
and the charm and bottom quark masses in the pole scheme.  Furthermore, we
also renormalize the wave function of the external bottom quark in the
on-shell scheme.  It is important to expand the bare two- and three-loop
correlators to ${\cal O}(\epsilon^2)$ and ${\cal O}(\epsilon)$, respectively,
in order to obtain the correct constant terms at NNLO.

In a second step we take into account the counterterms originating from
operator mixing (see Appendix~\ref{sec::OpMix} for details).  This requires
that in Eq.~(\ref{eqn:Gammatilde}) not only the physical operators $O_1$ and
$O_2$ are considered, but also evanescent operators~\cite{Buras:1989xd,Dugan:1990df}.

As shown in~\cite{Buras:1989xd}, Fierz symmetry can be restored order by 
order in perturbation theory by requiring that the anomalous dimension matrix (ADM) $\hat \gamma$,
which governs the renormalization group evolution of the Wilson coefficients $C_1$ and $C_2$, namely 
\begin{equation}
    \mu_b \frac{\mathrm{d} C_i}{\mathrm{d} \mu_b} =
    \gamma_{ji} C_j 
    \quad \text{for } i,j =1,2,
\end{equation}
fulfils
\begin{align}
    \gamma_{11} &= \gamma_{22},  &
    \gamma_{12} &= \gamma_{21}.
    \label{eqn:conditionsFierz}
\end{align}
We expand the ADM in series of $\alpha_s$:
\begin{equation}
    \hat \gamma =
    \hat \gamma^{(0)} 
    \frac{\alpha_s(\mu_b)}{4\pi} 
    +\hat \gamma^{(1)} 
    \left(\frac{\alpha_s(\mu_b)}{4\pi} \right)^2
    +\hat \gamma^{(2)} 
    \left(\frac{\alpha_s(\mu_b)}{4\pi} \right)^3
    +O(\alpha_s^4).
\end{equation}
The conditions~\eqref{eqn:conditionsFierz} can be imposed order by order in $\alpha_s$ 
by a proper definition of the evanescent operators.
At NLO, they are defined by~\cite{Buras:1989xd}:
\begin{align}
  E_1^{(1),q_1 q_2 q_3} &=
                          (\bar q_1^\alpha \gamma^{\mu_1 \mu_2 \mu_3} P_L b^\beta)
                          (\bar q_2^\beta
                          \gamma_{\mu_1 \mu_2 \mu_3} P_L
                          q_3^\alpha) - (16 - 4 \epsilon)O_1^{q_1 q_2 q_3},
  \notag \\
  E_2^{(1),q_1 q_2 q_3} &=
                          (\bar q_1^\alpha \gamma^{\mu_1 \mu_2 \mu_3} P_L b^\alpha)
                          (\bar q_2^\beta
                          \gamma_{\mu_1 \mu_2 \mu_3} P_L
                          q_3^\beta) - (16 - 4 \epsilon)O_2^{q_1 q_2 q_3},
\label{eqn:E1E2NLO}
\end{align}
where we introduced the notation
\begin{eqnarray}
    \gamma^{\mu_1 \dots \mu_N} = \gamma^{\mu_1} \dots \gamma^{\mu_N}.
\end{eqnarray}
While the $\epsilon$-independent term in front of $O_1$ and $O_2$ on the r.h.s.\ of~\eqref{eqn:E1E2NLO} is unique and obtained by requiring that 
the evanescent operator vanishes for $d=4$, the coefficients of $O(\epsilon)$ and higher
are in principle arbitrary.
This leads to the well-known scheme dependence of the ADM starting at NLO, which
eventually cancels for physical observables against the scheme dependence of the matching condition for the Wilson coefficients at the scale $\mu_W \simeq M_W$
and the matrix element of the effective operators at the low scale $\mu_b \sim m_b$.
The definition in Eq.~\eqref{eqn:E1E2NLO} leads to the NLO anomalous dimension matrix 
\begin{equation}
    \hat \gamma^{(1)} = 
    \begin{pmatrix}
        -\frac{21}{2}-\frac{2}{9}n_f & 
        \frac{7}{2}+\frac{2}{3}n_f \\[5pt]
        \frac{7}{2}+\frac{2}{3}n_f &
        -\frac{21}{2}-\frac{2}{9}n_f
    \end{pmatrix},
\end{equation}
which preserves Fierz symmetry up to $O(\alpha_s)$. 

At NNLO we have to consider the following evanescent operators:
\begin{align}
  E_1^{(1),q_1 q_2 q_3} &=
                          (\bar q_1^\alpha \gamma^{\mu_1 \mu_2 \mu_3} P_L b^\beta)
                          (\bar q_2^\beta
                          \gamma_{\mu_1 \mu_2 \mu_3} P_L
                          q_3^\alpha) - (16 - 4 \epsilon + A_2 \epsilon^2)O_1^{q_1 q_2 q_3},
  \notag \\
  E_2^{(1),q_1 q_2 q_3} &=
                          (\bar q_1^\alpha \gamma^{\mu_1 \mu_2 \mu_3} P_L b^\alpha)
                          (\bar q_2^\beta
                          \gamma_{\mu_1 \mu_2 \mu_3} P_L
                          q_3^\beta) - (16 - 4 \epsilon + A_2\epsilon^2)O_2^{q_1 q_2 q_3}, \notag\\
  E_1^{(2),q_1 q_2 q_3} &=
                          (\bar q_1^\alpha
                          \gamma^{\mu_1 \mu_2 \mu_3 \mu_4 \mu_5}
                          P_L b^\beta) (\bar q_2^\beta
                          \gamma_{\mu_1 \mu_2\mu_3\mu_4\mu_5}
                          P_L q_3^\alpha) -(256-224 \epsilon + B_1 \epsilon^2 )O_1^{q_1 q_2
                          q_3},  
  \notag \\
  E_2^{(2),q_1 q_2 q_3} &=
                          (\bar q_1^\alpha
                          \gamma^{\mu_1\mu_2\mu_3\mu_4\mu_5}
                          P_L b^\alpha) (\bar q_2^\beta
                          \gamma_{\mu_1\mu_2\mu_3\mu_4\mu_5}
                          P_L q_3^\beta)-(256-224 \epsilon + B_2 \epsilon^2 )O_2^{q_1 q_2 q_3}, 
  \label{eqn:evaoperators}
\end{align}
At $O(\alpha_s^2)$ also $O(\epsilon^2)$ terms must be considered in the coefficients multiplying
$O_1^{q_1 q_2 q_3}$ and $O_2^{q_1 q_2 q_3}$ in the evanescent operator definitions.
By imposing that also the NNLO ADM $\gamma^{(2)}$ fulfils Eq.~\eqref{eqn:conditionsFierz},
we can fix the coefficient $A_{2}$, $B_1$ and $B_2$.
To this end, we take the expressions for $\gamma^{(2)}$ in the CMM basis~\cite{Gorbahn:2004my}
and perform a transformation to the historical basis. For simplicity we restrict ourself to $O_1$ and $O_2$,
neglecting the penguin operators. We give more details on the basis transformation in the Appendix~\ref{sec::OpMix}.
We find that the Fierz symmetry is preserved by a one-parameter class of renormalization scheme 
defined by:
\begin{align}
  B_1 &= -\frac{4384}{115} 
  - \frac{32}{5} n_f + 
  A_2 \Bigg( \frac{10388}{115} -  \frac{8}{5} n_f \Bigg), \notag \\
  B_2 &= -\frac{38944}{115} - \frac{32}{5} n_f + 
  A_2 \Bigg( \frac{19028}{115} - \frac{8}{5} n_f \Bigg).
  \label{eqn:solutionB1B2}
\end{align}
We highlight two notable scheme choices. The first one is
\begin{align}
    A_2 &= - 4,  &
    B_1 &= -\frac{45936}{115} , &
    B_2 &= -\frac{115056}{115},
    \label{eqn:standard_evanescent_definition}
\end{align}
which gives a definition of $E_{1,2}^{(2)}$ independent on $n_f$ however with $B_1 \neq B_2$. 
We adopt the values in Eq.~\eqref{eqn:standard_evanescent_definition} as reference scheme for the evanescent operators. 
All scheme dependent quantities will be given relative to this choice.
%
Another notable choice would be
\begin{align}
    A_2 &= + 4, &
    B_1 &= B_2 = \frac{1616}{5} - \frac{64}{5}n_f,
\end{align}
which leads to the same coefficient for $E_{1}^{(2)}$ and $E_{2}^{(2)}$ at $O(\epsilon^2)$.
In principle, we could have defined the evanescent operators $E_{1}^{(1)}$ and $E_{2}^{(1)}$
in a more general way with unequal coefficients at order $\epsilon^2$, namely with $A_1 \neq A_2$. 
In this case one would find a class of renormalization schemes governed by two parameters 
instead of one. However we observe that the only solution
independent on $n_f$ still correspond to the case given in Eq.~\eqref{eqn:solutionB1B2},
therefore for simplicity we set $A_1 = A_2$, and verify that the dependence on 
$A_2$ drops out in the total width.

With the evanescent operator definition in Eq.~\eqref{eqn:evaoperators}, we 
compute at LO all correlators which involve
$\{O_1,O_2,E_1^{(1)},E_2^{(1)},E_1^{(2)},E_2^{(2)}\}$ and at NLO those with
$\{O_1,O_2,E_1^{(1)},E_2^{(1)}\}$. At NNLO only the physical operators have to
be considered. 
We compute the renormalization constants of the operators up to $O(\alpha_s^2)$ 
and check that after a basis transformation we reproduce the known results in the CMM basis.
Further details on the calculation of the renormalization constants, and their explicit expressions,
are given in Appendix~\ref{sec::OpMix}.

Once all renormalization constants are taken into account, we arrive at
a finite expression for the decay rate up to NNLO. At this point it
is straightforward to choose different renormalization schemes
for the quark masses.

One important cross check on the finite result is to verify that in the
massless limit the coefficients in front of $C_1^2$ and $C_2^2$ are equal
up to $O(\alpha_s^2)$. This is a consequence of Fierz symmetry for the operators
$O_1$ and $O_2$, whose contributions to the rate become indistinguishable if all final-state quarks are massless.
Notice that this is a necessary but not sufficient condition for imposing Fierz symmetry
in the renormalized results.



\section{Results for the total rates}
\label{sec:results}

We present in this section our results for the squared amplitude up to $O(\alpha_s^2)$ 
and combine them with the perturbative expansion of the Wilson coefficient at the scale $\mu_b \sim m_b$
up to NNLO.
Let us write the decay rates in the following way:
\begin{equation}
    \Gamma^{q_1q_2q_3} = 
    \Gamma_0
    \Bigg[
    C_1^2(\mu_b)
    G_{11}^{q_1q_2q_3} 
   + C_1(\mu_b)
    C_2(\mu_b)
    G_{12}^{q_1q_2q_3} 
    + C_2^2(\mu_b)
    G_{22}^{q_1q_2q_3} 
    \Bigg],
    \label{eqn:decayrateformula}
\end{equation}
where  $\Gamma_0 = G_F^2m_b^5 |\lambda_{q_1q_2q_3}|^2 /(192 \pi^3)$. For the sake of clarity, we omit in the following the flavour indices for $G_{ij}$. They can always be reconstructed from the context in which these quantities are used.
The functions $G_{ij}$ are the interference terms between the insertion of the operators $O_i$ and $O_j$.
They depend on the mass ratio $\rho$ and the renormalization scale $\mu_b$. 
Their perturbative expansion in $\alpha_s$ is given by
\begin{equation}
    G_{ij}= 
    G_{ij}^{(0)}
    +\frac{\alpha_s}{\pi} G_{ij}^{(1)}
    +\left(\frac{\alpha_s}{\pi} \right)^2 G_{ij}^{(2)}
    +O(\alpha_s^3),
\end{equation}
where $\alpha_s \equiv \alpha_s^{(5)}(\mu_b)$ is the strong coupling constant with $n_f=5$ active quarks at the renormalization scale $\mu_b$. 

\subsection{\boldmath One massive quark: $b \to c\bar u d$}
We report here the analytic expressions at LO and NLO (for $\mu_b = m_b$) of the interference term
for the decays with one massive charm quark in the final state: $b \to c \bar u d$ and $b \to c \bar u s$.
Their analytic expressions written in terms of HPLs read
\begin{equation}
    G_{11}^{(0)} = G_{22}^{(0)} = 
    \frac{3}{2} G_{12}^{(0)} = 3\Big(1 - 8 \rho^2 + 8 \rho^6 - \rho^8 - 24 \rho^4 H_0( \rho ) \Big).
    \label{eq::G_b2cud_LO}
\end{equation}
\begin{align}
    G_{11}^{(1)} &=
    \frac{31}{2}-\frac{554 \rho ^2}{3}+\frac{554 \rho ^6}{3}-\frac{31 \rho ^8}{2}
    +\pi ^2 \left(-2+16 \rho ^2+48 \rho ^4-\frac{16 \rho ^6}{3}+\frac{2 \rho ^8}{3}\right)
    \notag \\ &
    +\left(-\frac{62}{3}+\frac{640 \rho ^2}{3}-\frac{640 \rho ^6}{3}+\frac{62 \rho ^8}{3}\right) H_{-1}(\rho )
    +\left(\frac{62}{3}-\frac{640 \rho ^2}{3}+\frac{640 \rho ^6}{3}-\frac{62 \rho ^8}{3}\right) H_1(\rho )
    \notag \\ &
    +\left(-96 \rho ^2-120 \rho ^4+32 \pi ^2 \rho ^4+\frac{992 \rho ^6}{3}-\frac{124 \rho ^8}{3}\right) H_0(\rho )
    -\left(288 \rho ^4+64 \rho ^6-8 \rho ^8\right)  \left(H_0(\rho )\right)^2
    \notag \\ &
    +\left(8-64 \rho ^2-576  \rho ^4-64 \rho ^6+8 \rho ^8\right) \Big[H_{0,1}(\rho )-H_{0,-1}(\rho )\Big] \, ,
    \\ 
    G_{22}^{(1)} &=
    \frac{31}{2}-\frac{550 \rho ^2}{3}+\frac{550 \rho ^6}{3}-\frac{31 \rho ^8}{2}
    +\pi ^2 \left(-2+64 \rho ^3-32 \rho ^4+64 \rho ^5-2 \rho ^8\right)
    \notag \\ &
    +\left(-288 \rho ^4-8 \rho ^8\right) \left(H_0(\rho )\right)^2
    +\left(-\frac{34}{3}+\frac{128 \rho ^2}{3}-\frac{128 \rho ^6}{3}+\frac{34 \rho ^8}{3} \right) H_{-1}(\rho )   
    \notag \\ &
    +\left(\frac{34}{3}-\frac{128 \rho   ^2}{3}+\frac{128 \rho ^6}{3}-\frac{34 \rho ^8}{3}\right) H_1(\rho )
    +\left(-80 \rho ^2-432 \rho ^4+\frac{16 \rho ^6}{3}-\frac{68 \rho ^8}{3}\right) H_0(\rho ) 
    \notag \\ &
    +\left( 16+256 \rho ^3+480 \rho ^4+256 \rho ^5+16 \rho ^8\right) H_0(\rho ) H_{-1}(\rho )   
    -\Big(16-256 \rho ^3+480 \rho ^4-256 \rho ^5 
    \notag \\ &
    +16 \rho ^8\Big) H_0(\rho )  H_1(\rho )
    -\left(24+256 \rho ^3+384 \rho ^4+256 \rho ^5+24 \rho ^8\right) H_{0,-1}(\rho )
    \notag \\ &
    +\left(24-256 \rho ^3+384 \rho ^4-256 \rho ^5+24 \rho ^8\right) H_{0,1}(\rho ) \, ,
    \\
    G_{12}^{(1)} &=
    -17+\frac{1828 \rho ^2}{9}-\frac{1828 \rho ^6}{9}+17 \rho ^8
    -\pi ^2 \bigg(\frac{4}{3}+\frac{32 \rho ^2}{3}-\frac{128 \rho ^3}{3}+\frac{160 \rho ^4}{3}-\frac{128 \rho ^5}{3}+\frac{32 \rho ^6}{3}
    \notag \\ &
    +\frac{4 \rho  ^8}{3}\bigg)
   -\left(\frac{160 \rho ^2}{3}-352 \rho ^4+\frac{1504 \rho ^6}{9}+\frac{376 \rho ^8}{9} \right) H_0(\rho ) 
   +\bigg(\frac{116}{9}+\frac{1088 \rho ^2}{9}-\frac{1088 \rho ^6}{9}
   \notag \\ &
   -\frac{116 \rho   ^8}{9}\bigg) H_1(\rho )
   +\left(-\frac{116}{9}-\frac{1088 \rho ^2}{9}+\frac{1088 \rho ^6}{9}+\frac{116 \rho   ^8}{9} \right) H_{-1}(\rho ) 
   +\bigg(\frac{32}{3}+\frac{512 \rho ^3}{3}
   \notag \\ &
   +320 \rho ^4+\frac{512 \rho ^5}{3}+\frac{32 \rho ^8}{3}\bigg) H_0(\rho ) H_{-1}(\rho ) 
   -\left(192 \rho ^4-128 \rho ^6+\frac{16 \rho ^8}{3}\right) \left(H_0(\rho )\right)^2
   \notag \\ &
   +\left(-\frac{32}{3}+\frac{512 \rho ^3}{3}-320 \rho ^4+\frac{512 \rho ^5}{3}-\frac{32 \rho   ^8}{3}\right) H_1(\rho ) H_0(\rho ) 
   +\bigg(-16-128 \rho ^2-\frac{512 \rho ^3}{3}
   \notag \\ &
   -640 \rho ^4-\frac{512 \rho ^5}{3}-128 \rho ^6-16 \rho ^8\bigg) H_{0,-1}(\rho )
   +\bigg(16+128 \rho ^2-\frac{512 \rho ^3}{3}+640 \rho^4-\frac{512 \rho ^5}{3}
   \notag \\ &
   +128 \rho ^6+16 \rho ^8\bigg) H_{0,1}(\rho ) \, .
\end{align}
The HPLs are defined by 
\begin{align}
    H_{w_1,\vec{w}}(\rho) &= \int\limits_{0}^{\rho} 
    {\rm dt} f_{w_1}(t) \, H_{\vec{w}}(t) ~,
\end{align}
with the letters 
\begin{align} 
    f_{0}(t) &= \frac{1}{t}, &
    f_{1}(t) &= \frac{1}{1-t}, &
    f_{-1}(t) &= \frac{1}{1+t}~,
\end{align}
and the regularization $H_{0}(t) = \log(t)$.
The functions needed at NLO can also be expressed in terms of classical
logarithms and polylogarithms:
\begin{align*}
    H_{0}(\rho) &= \log(\rho), &
    H_{1}(\rho) &= - \log(1-\rho), &
    H_{-1}(\rho) &= \log(1+\rho) , \\
    H_{0,1}(\rho) &= \text{Li}_2(\rho) , &
    H_{0,-1}(\rho) &= -\text{Li}_2(-\rho) ~.
\end{align*}
In the massless limit, the interference terms at NLO reduce to\footnote{Our coefficient $G_{12}^{(1)}$ 
differs from the result usually quoted from~\cite{Altarelli:1991dx,Bagan:1994zd}. 
In these articles part of the $O(\alpha_s)$ correction to the Wilson coefficients is reabsorbed into the 
interference terms, however we prefer to keep the two corrections separated since they have different origin.}
\begin{align}
    G_{11}^{(1)} &= G_{22}^{(1)}  = 
    \frac{31}{2}-2 \pi ^2, &
    G_{12}^{(1)} &= -17-\frac{4 \pi ^2}{3}.
    \label{eqn:NLOmassless}
\end{align}
For illustration we present in the following our numerical results for the NNLO interference terms $G_{ij}^{(2)}$ 
as a series expansion around $\rho=0$ up to $\rho^7$. 
For the numerical evaluation it is more convenient to use the expansions close to the physical value of $m_c/m_b$.
For the colour factors, we insert their values in QCD and set $n_l=3, n_c =1$ and $n_b=1$,
where $n_l$ denotes the contribution from closed massless fermion loops while the $n_c$ and $n_b$
contributions arise from closed fermion loops with masses $m_c$ and $m_b$, respectively.
Our results read:
\begin{align}
    G_{11}^{(2)} & =
    13.4947
    -24.6740 \rho 
    +\left(115.542-625.679 \lr+8 \lr^2\right) \rho ^2
    +(-76.2198+210.552 \lr) \rho ^3
    \notag \\ &
    +\left(1829.82+2820.10 \lr-1058.37 \lr^2+32 \lr^3\right) \rho ^4
    +(-74.0184+574.630 \lr) \rho ^5
    \notag \\ &
    +\left(-2197.51-12.1531 \lr+892.933 \lr^2-257.185 \lr^3\right) \rho ^6
    \notag \\ &
    +(-371.871+433.678 \lr) \rho ^7
    +O(\rho^8), 
    \label{eq::G11}\\[5pt]
    G_{22}^{(2)} &=
    13.4947
    -24.6740 \rho 
    +\left(-2647.00-1257.33 \lr+8\lr^2\right) \rho ^2
    +(-2883.63-3842.57 \lr) \rho ^3
    \notag \\ &
    +\left(3403.78-4764.59 \lr-777.272 \lr^2+32\lr^3\right) \rho ^4
    +(-601.507-2397.66 \lr) \rho ^5
    \notag \\ &
    +\left(2833.10-559.776 \lr+264.958 \lr^2-21.3333 \lr^3\right) \rho ^6
    \notag \\ &
    +(-440.110+57.8599 \lr) \rho ^7
    +O(\rho^8), 
    \label{eq::G12}\\[5pt]
    G_{12}^{(2)} &=
    -72.8420
    -16.4493 \rho 
    +\left(-279.953+63.3413 \lr+5.33333 \lr^2\right) \rho ^2
    \notag \\ &
    +(-3707.85-2702.08 \lr) \rho ^3
    +\left(2164.12-2197.00 \lr+2041.82 \lr^2+21.3333 \lr^3\right) \rho ^4
    \notag \\ &
    +(-888.121-1177.33 \lr) \rho ^5
    +\left(1987.97-4886.67 \lr-174.131 \lr^2+632.889 \lr^3\right) \rho ^6
    \notag \\ &
    +(846.185-66.7025 \lr) \rho ^7
        +O(\rho^8), \label{eq::G22}
\end{align}
where $\lr = \log(\rho)$.
In Eqs.~(\ref{eq::G11}) to~(\ref{eq::G22}) we present six significant
digits for the numerical coefficients
and suppress tailing zeros. Note that in most cases we have a higher accuracy.

\begin{figure}
    \centering
         \includegraphics[width=0.47\textwidth]{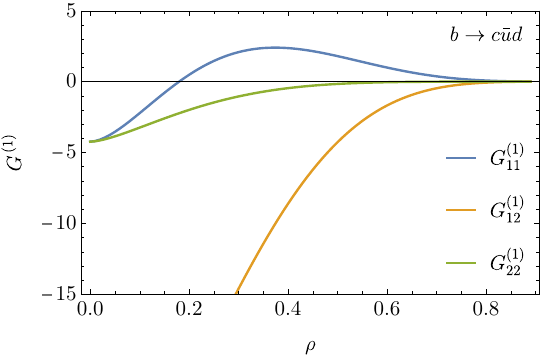}
         \hfill
         \includegraphics[width=0.47\textwidth]{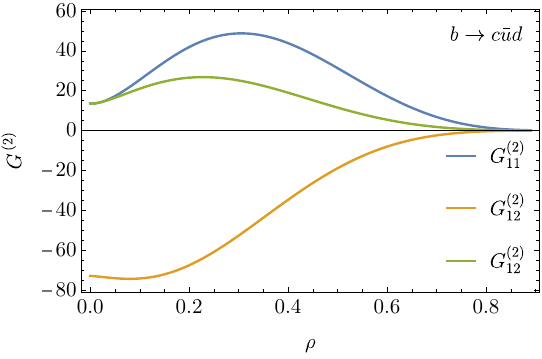}
    \caption{The NLO (left) and NNLO (right) corrections to different combinations of 
    Wilson coefficients for $b \to c \bar u d$ as defined in Eq.~\eqref{eqn:decayrateformula} as
    functions of $\rho = m_c/m_b$ for $\mu_b = m_b$.}
    \label{fig:plotsGij}
\end{figure}

In Fig.~\ref{fig:plotsGij} we show our predictions for the interference terms $G_{11}^{(n)}$, $G_{12}^{(n)}$ and $G_{22}^{(n)}$ at NLO ($n=1$) and NNLO ($n=2$) as a function of $\rho = m_c/m_b$  with $\mu_b = m_b$.
Both at NLO and NNLO, the limit $\rho \to 0$ is finite after renormalization, as expected from the cancellation of
mass singularities in inclusive observables. To this end, as discussed for the semileptonic decay in~\cite{Egner:2023kxw}, it is crucial to properly treat the master integrals around the singular point of the differential equations $\rho=1/3$. In particular, one has to take into account that there are master integrals which have no imaginary part for $\rho>1/3$.
Finally we note that $G_{11}$ and $G_{22}$ have the same limit for $\rho \to 0$ both at NLO and NNLO as required by Fierz symmetry.
We remind that the expressions given for $G_{ij}^{(1)}$, $G_{ij}^{(2)}$ and the plots in Fig.~\ref{fig:plotsGij}
are scheme dependent, relative to our default choice of the evanescent operators in Eqs.~\eqref{eqn:evaoperators} and~\eqref{eqn:standard_evanescent_definition}.

We can now turn to the prediction for the total rate $\Gamma^{cdu}$. 
We have to combine our results for the interference terms up to NNLO and 
the perturbative expansion of the Wilson coefficients at the low scale $\mu_b$ obtained 
with RGE at NNLO.
The matching conditions for $C_1$ and $C_2$ in the historical basis 
can be obtained starting from those in the CMM basis~\cite{Bobeth:1999mk} 
and then performing a basis transformation (see Appendix~\ref{sub::basis}, in particular Eq.~\eqref{eqn:basis_change_Wilsoncoefficients}).
The explicit expressions in the historical basis with anticommuting $\gamma_5$ read \cite{Gorbahn:2004my}:
\begin{align}
    C_1(\mu_W) &=
    \frac{\alpha_s}{4 \pi}
    \left(\frac{11}{2}+3L\right)
    \notag \\ &
    +\left(\frac{\alpha_s}{4 \pi}\right)^2
    \left[
    \frac{64385}{3312}
    +\frac{17 \pi^2}{6}
    -\frac{5}{9}n_f
    +\frac{415 }{12} L
    +\frac{17}{2} L^2
    -\frac{1}{2} T\left(\frac{m_t^2}{M_W^2}\right)
    \right]
    +O(\alpha_s^3),
    \notag \\
    C_2(\mu_W) &=
    1
    -\frac{\alpha_s}{4 \pi} \left(\frac{11}{6}+L\right)
    \notag \\ &
    +\left(\frac{\alpha_s}{4 \pi} \right)^2 
    \left[
    -\frac{3894253}{49680}
    +\frac{7 \pi ^2}{18}
    -\frac{55}{36} L
    +\frac{7}{6} L^2
    +\frac{5}{27} n_f
    +\frac{1}{6} T\left(\frac{m_t^2}{M_W^2}\right)
    \right]
    +O(\alpha_s^3),
\end{align}
where $L=\log(\mu_W^2/M_W^2)$. The function $T(x)$ parametrizes the top-quark mass effects. 
Its explicit expression can be retrieved from Eq.~(19) in Ref.~\cite{Bobeth:1999mk}.

The solution of the RGE for the Wilson coefficients up to NNLO is well known and 
presented in Refs.~\cite{Buras:1991jm,Gorbahn:2004my}.
It requires the NNLO ADM in the historical basis, with the evanescent operator definition in Eq.~\eqref{eqn:evaoperators}.
After performing a basis transformation of $\gamma^{(2)}$ from the CMM basis to the historical basis we obtain:
{\footnotesize
\begin{equation}
\hat \gamma^{(2)} =
\begin{pmatrix}
 \frac{1340209}{460}-n_f \left(\frac{1190291}{6210}-\frac{80 \zeta_3}{3}\right) +\frac{130 }{81} n_f^2& 
 \frac{401635}{276}-672 \zeta_3-n_f \left(\frac{16657}{414}+80 \zeta_3\right) -\frac{130 }{27} n_f^2\\[10pt]
 \frac{401635}{276}-672 \zeta_3-n_f \left(\frac{16657}{414}+80 \zeta_3\right) -\frac{130 }{27} n_f^2&
 \frac{1340209}{460}-n_f \left(\frac{1190291}{6210}-\frac{80 \zeta_3}{3}\right) +\frac{130 }{81} n_f^2 
\end{pmatrix},
\label{eqn:gamma2}
\end{equation}
}

With the matching conditions and the ADM up to NNLO, we can 
calculate\footnote{We perform the running from $\mu_W$ to $\mu_b$
  both in the historical 
  and, as a cross check, in the CMM basis.  
  In the latter case we apply the basis change  relations from
  Appendix~\ref{sub::basis}.}
the values of the 
Wilson coefficients at the low-energy scale $\mu_b \sim m_b$
\begin{equation}
    C_i(\mu_b) = 
    C_i^{(0)}(\mu_b) 
    +\frac{\alpha_s(\mu_b)}{4 \pi}C_i^{(1)}(\mu_b) 
    +\left(\frac{\alpha_s(\mu_b)}{4 \pi} \right)^2C_i^{(2)}(\mu_b) 
    +O(\alpha_s^3),
\end{equation}
which is appropriate for studying nonleptonic decay. 
We report in Tab.~\ref{tab:valuesC1C2} the values for the Wilson coefficients at the reference scale
$\mu_b = 4.7$~GeV and matching scale $\mu_W=M_W$. For the numerical evaluation of $\alpha_s(\mu_b)$ we use
the five-loop RGE implemented in \texttt{RunDec}~\cite{Chetyrkin:2000yt,Herren:2017osy} with $\alpha_s(M_Z) = 0.1179$.
\begin{table}[]
    \centering
    \begin{tabular}{c|rr}
    & $i=1$ & $i=2$ \\
    \hline
    $C^{(0)}_i(\mu_b)$ & $-$0.2511 & 1.109 \\
    $C^{(1)}_i(\mu_b)$ &  4.382 & $-$2.016 \\
    $C^{(2)}_i(\mu_b)$ &  36.63 & $-$82.19 \\
    \end{tabular}
    \caption{Values of the Wilson coefficients $C_1(\mu_b)$ and $C_2(\mu_b)$
    in the historical basis at LO, NLO and NNLO  at the scale $\mu_b = 4.7$~GeV.
    The matching scale is $\mu_W = M_W$.}
    \label{tab:valuesC1C2}
\end{table}
After inserting the values for $C_1(\mu_b)$ and $C_2(\mu_b)$ into Eq.~\eqref{eqn:decayrateformula} and re-expanding it in series of $\alpha_s$ 
we obtain the perturbative expansion for the branching ratio in the on-shell scheme
\begin{align}
    \Gamma^{cdu} &= \Gamma_0 
    \left[ 1.89907 
    +1.77538 \frac{\alpha_s}{\pi}
    +14.1081 \left(\frac{\alpha_s}{\pi}\right)^2 
    \right]
   ,
    \label{eqn:Gamma_numeric} 
\end{align}
with $\alpha_s = \alpha_s^{(5)}(m_b)$, $m_b^{\mathrm{OS}} = 4.7$~GeV and $m_c^{\mathrm{OS}} = 1.3$~GeV. 
As a cross check, we repeated the calculation of the interference terms,
the ADM and the Wilson coefficients at NNLO without specifying the numerical value for $A_2$ in Eq.~\eqref{eqn:evaoperators} and~\eqref{eqn:solutionB1B2}. 
We explicitly verified that $A_2$ drops out in Eq.~\eqref{eqn:Gamma_numeric},
so that the rate is independent on the scheme adopted for the evanescent operators. 
This represent a strong validation of our computational setup.

\begin{figure}
    \centering
\includegraphics[width=0.8\textwidth]{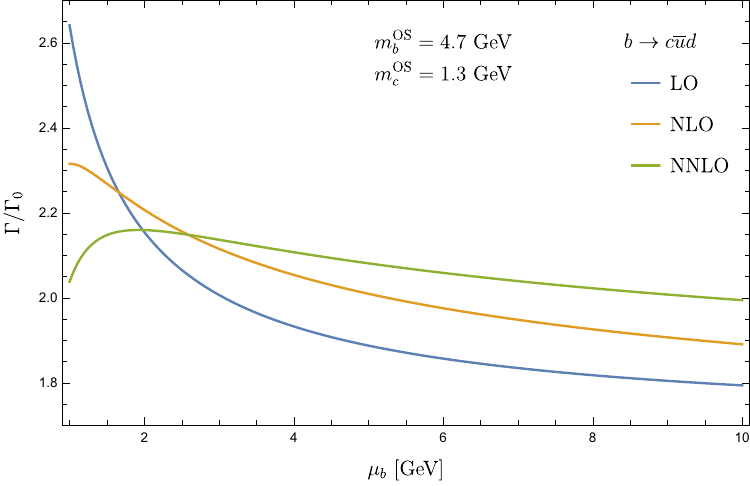}
    \caption{The dependence of the rate for $b \to c \bar u d$ on the renormalization 
    scale $\mu_b \sim m_b$ at LO, NLO and NNLO in the on-shell scheme.}
    \label{fig:mus_dependence}
\end{figure}

The dependence of the rate on the renormalization scale is presented in Fig.~\ref{fig:mus_dependence}. 
In the plot we vary the scale $\mu_b$ associated to
the strong coupling constant and the Wilson coefficients from $1\, \mathrm{GeV} < \mu_b < 10 \, \mathrm{GeV}$.
We estimate the theoretical uncertainty by determining maximum and minimum for 
$\mu_b\in\{m_b/2,2m_b\}$ and dividing the result by two.
We observe that the scale uncertainty is significantly reduced
once higher order QCD corrections are included. 
Whereas at leading order the scale variation between $m_b/2$ and $2 m_b$ yields a relative uncertainty 
of about 7\%, which reduces to 6.3\% at NLO, the inclusion of the NNLO corrections 
further reduce the scale uncertainty to 3.5\% relative to the central value at $\mu_b=m_b$.
At the central scale $\mu_b=m_b$ the $O(\alpha_s)$ corrections are about 6.5\% of the LO result
and the $O(\alpha_s^2)$ corrections are less than 3.5\% of the prediction at NLO. 
Note that close to $\mu_b=m_b/2$ the NNLO corrections vanish.

\subsection{Massless contribution and secondary charm pair production}

\begin{figure}
    \centering
\includegraphics[width=0.8\textwidth]{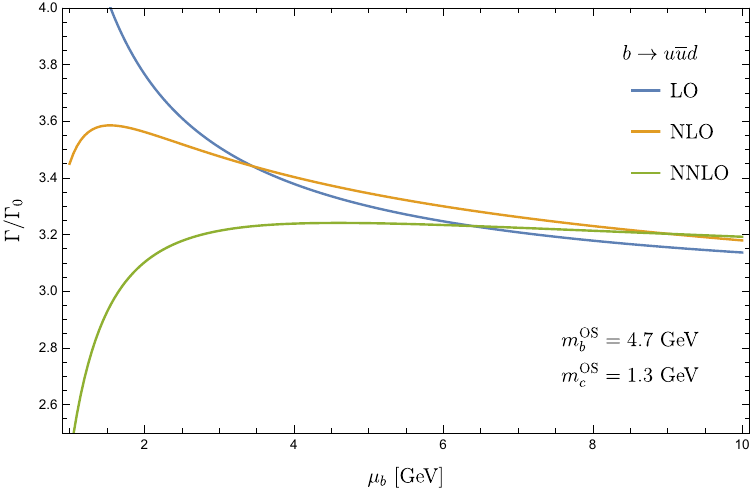}
    \caption{The dependence of the rate for $b \to u \bar u d$ on the renormalization 
    scale $\mu_b \sim m_b$ at LO, NLO and NNLO in the on-shell scheme.}
    \label{fig:mus_dependence_uud}
\end{figure}

From the expressions for the $b\to c\bar ud$ decay, it is possible to take the limit $\rho \to 0$ and
obtain the decay rate for $b \to u\bar u d $. We then add the $U_c$ contribution 
arising at $O(\alpha_s^2)$ from the insertion of a closed charm loop into 
the gluon propagators (see for instance the diagram in Fig.~\ref{fig:NLOnnlep}(i).
After inserting the values for $C_1$ and $C_2$,
we obtain the perturbative expansion for the branching
ratio in the massless limit:
\begin{align}
    \Gamma^{ud u} &= \Gamma_0 
    \left[ 3.31981 
    +0.597456 \frac{\alpha_s}{\pi}
    +{ \left( -25.645+0.254978_{U_c} \right) }\left(\frac{\alpha_s}{\pi}\right)^2 
    \right] ,
\end{align}
where the last term arises from the $U_c$ contribution and 
depends on the ratio between the bottom and the charm mass.
We use $m_b^{\mathrm{OS}} = 4.7$~GeV and $m_c^{\mathrm{OS}} = 1.3$~GeV.
The dependence of the rate on the renormalization scale $\mu_b$ for the massless decay
$b\to u\bar ud$ is presented in Fig.~\ref{fig:mus_dependence_uud}. 
We observe also here that  for $m_b/2 < \mu_b < 2m_b$ the scale uncertainty is reduced from a relative 7\% at LO, to 5\% at NLO, to less than 1.3\%
after incorporating the NNLO corrections.
At the central scale the NLO corrections are positive and amount to about 2\% whereas 
the NNLO corrections are approximately twice as big and negative. However, close to $\mu_b=2m_b$ the NNLO corrections vanish.

\subsection{Two massive charm quarks: \boldmath $b \to c\bar c s$}

Also for the channel with two massive charm quarks in the final
state we obtain analytic results at LO and NLO. For convenience we
present in the following expansions 
around $\rho=0$. At LO we obtain
\begin{align}
    G_{11}^{(0)} &= G_{22}^{(0)} =
    \frac{3}{2} G_{12}^{(0)} 
    \notag \\ &
    =\frac{3 (1-x^2)\left(x^{12}-8 x^{10}-43 x^8-80 x^6-43   x^4-8 x^2+1\right)}{\left(x^2+1\right)^7}
    \notag \\ &
    -\frac{144 \left(x^2-x+1\right) \left(x^2+x+1\right) \left(x^4+3 x^2+1\right) x^4 \log (x)}{\left(x^2+1\right)^8}
    \notag \\ &
    =3\Big(1 - 16 \rho^2 + 24 \rho^4 - 32 \rho^6 +2 \rho^8 + 32 \rho^{10} + \left( - 48 \rho^4 + 48 \rho^8 \right) H_0( \rho ) \Big) + \mathcal{O}\left( \rho^{12} \right)
    \,.
\end{align}

The analytic expressions 
for $b\to c \bar c s$ at NLO are rather lengthy and we
report them only in the ancillary files attached to this paper~\cite{progdata,egner_2024_11639756}.
In the following we provide only the first few terms in an expansion
around $\rho=0$.
\begin{align}
    G_{11}^{(1)} &= 
    \frac{31}{2}-2 \pi ^2
    -\left(320-32 \pi ^2+ 192 l_\rho\right) \rho ^2
    +\left[
    -276+64 \pi ^2
    +\left(336+64 \pi ^2\right) l_\rho
    -576 l_\rho^2
    \right]\rho ^4
    \notag \\ &
    +\left(
    -\frac{1552}{3}
    +\frac{160 \pi ^2}{3}
    -\frac{3200 l_\rho}{3}
    -128 l_\rho^2
    \right) \rho ^6
    +\left[
    -\frac{7}{2}
    -108 \pi ^2
    +288 \zeta (3)
    \right.
    \notag \\ &
    \left.
    + \left(\frac{88}{3}-64 \pi ^2\right) l_\rho
    +1264 l_\rho^2
    \right]\rho ^8 +O(\rho^{9}),
    \\
    G_{22}^{(1)} &= 
    \frac{31}{2}-2 \pi ^2
    +\left(-320+16 \pi ^2-192 l_\rho\right) \rho ^2+
    64 \pi ^2 \rho ^3
    +\left[-636-576 l_\rho^2+16 \pi ^2
    \right.
    \notag \\ &
    \left.
    +l_\rho \left(528+32 \pi ^2\right)\right]\rho ^4
    +\left(\frac{2696}{9}-\frac{16 \pi ^2}{3}
    -\frac{5312 l_\rho}{3}-64 l_\rho^2\right) \rho ^6
    -64 \pi ^2 \rho ^7
    \notag \\ &
    +\left[-\frac{14236}{75}
    -68 \pi ^2
    -288 \zeta (3)
    +1440 l_\rho^2
    +l_\rho \left(-\frac{628}{5}+32 \pi ^2\right)
    \right]\rho ^8 
    +O(\rho^9) ,\\
    G_{12}^{(1)} &=
    -17-\frac{4 \pi ^2}{3}
    +\left(336
    +16 \pi ^2
    -128 l_\rho
    \right) \rho ^2
    -\frac{128 \pi ^2}{3} \rho^3
    +\left[
    -328
    -384 l_\rho^2
    +l_\rho \left(1120
    \right. \right . 
    \notag \\ &
    \left. \left .
    +\frac{128 \pi ^2}{3}\right)\right] \rho ^4
    +\frac{256 \pi ^2 }{3}\rho^5
    +\left[
    \frac{6008}{27}
    +\frac{176 \pi ^2}{9}
    +192 \zeta (3)
    +l_\rho \left(\frac{128}{9}-\frac{128 \pi ^2}{3}\right)
    +64 l_\rho^2
    \right]\rho ^6 
    \notag \\ &
     -128 \pi ^2 \rho ^7
    +\left[
    \frac{237341}{675}
    -\frac{104 \pi ^2}{9}
    +192 \zeta (3)
    +l_\rho \left(-\frac{47152}{45}-\frac{128 \pi ^2}{3}\right)
    +\frac{1376 l_\rho^2}{3}
    \right]
    \rho^8 
    \notag \\ &
    + O(\rho^9).
\end{align}
One observes that the result at $\rho=0$ coincides with the massless limit in Eq.~\eqref{eqn:NLOmassless}.
For the NNLO result we obtain expansions around $\rho=0$, $\rho=1/5$ and $\rho=1/3$ 
using ``expand and match'' as described above. The expressions for the NNLO result expanded around $\rho=0$ read
\begin{align}
    G_{11}^{(2)} &= 13.4947 -
 24.6740\rho+
 \left(-533.154 - 1251.35l_\rho + 16.0000l_\rho^2\right)\rho^2 \nonumber\\
 &+ 
 \left(2998.33+ 210.551l_\rho \right)\rho^3 +
 \left(116.662 + 10686.6l_\rho - 628.278l_\rho^2 
 + 64.0000l_\rho^3\right) \rho^4 \nonumber \\
 & +
 \left(-2364.37 + 903.617l_\rho \right) \rho^5 +
 \left(5409.89 - 6392.16l_\rho - 5833.18l_\rho^2 - 507.259l_\rho^3 \right) \rho^6 \nonumber\\
 &+
\left(-135.120 + 3582.30l_\rho \right)\rho^7+
 \left(-2320.40 -15254.6l_\rho + 6047.29l_\rho^2 +2654.91l_\rho^3 \right)\rho^8, \\
   G_{22}^{(2)} &= 13.4947 - 24.6740\rho    + \left(-3295.69 -1883.01l_\rho +16.0000l_\rho^2 \right) \rho^2 \nonumber\\
 &+ \left(190.918 -3842.56l_\rho \right) \rho^3 + \left(4770.24 +5244.57l_\rho -347.178l_\rho^2 +64.0000l_\rho^3\right) \rho^4 \nonumber\\
 &   + \left(1299.89 - 542.170l_\rho\right) \rho^5   + \left(5832.60 -7262.42l_\rho -10760.9l_\rho^2 -271.407l_\rho^3 \right) \rho^6 \nonumber\\
 &   + \left(1667.93 + 1806.31l_\rho \right) \rho^7
   + \left(-30465.4 -19154.5l_\rho +9435.28l_\rho^2 +4491.90l_\rho^3 \right)\rho^8 ,\\
   G_{12}^{(2)} &= -72.8419- 16.4493\rho 
   + \left(3160.66 + 1152.54l_\rho + 10.6666l_\rho^2\right) \rho^2\nonumber\\
 &   + \left(3980.72 +2702.07l_\rho\right) \rho^3
   + \left(-5915.68 + 8133.91l_\rho + 3393.24l_\rho^2 +42.6666l_\rho^3 \right)\rho^4 \nonumber\\
 &   + \left(-8194.22 +5070.78l_\rho \right) \rho^5
   + \left(18013.7 +3710.59l_\rho +1222.25l_\rho^2 +278.518l_\rho^3 \right)\rho^6\nonumber\\
 &   + \left(184.414 + 7746.34l_\rho \right) \rho^7
   + \left(7997.58+ 2111.25l_\rho - 9970.14l_\rho^2 - 773.168l_\rho^3 \right)\rho^8 .
\end{align}

\begin{figure}[t]
    \centering
         \includegraphics[width=0.47\textwidth]{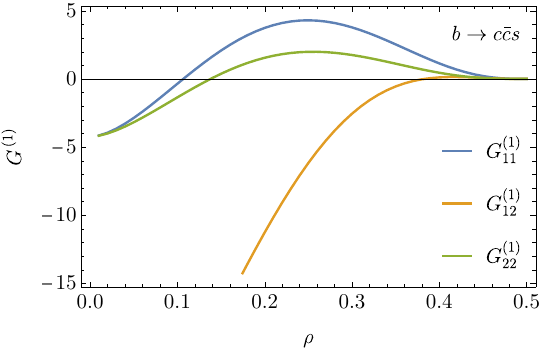}
         \hfill
         \includegraphics[width=0.47\textwidth]{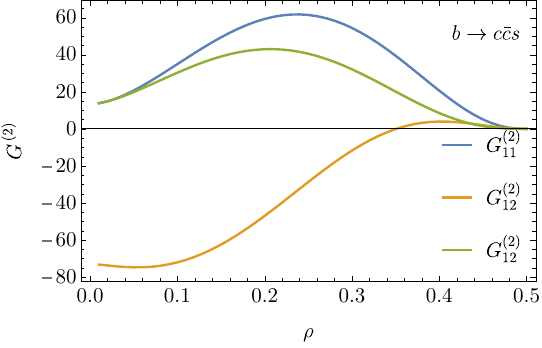}
    \caption{The NLO (left) and NNLO (right) corrections to different combinations of 
    Wilson coefficients for $b \to c \bar c s$ as defined in Eq.~\eqref{eqn:decayrateformula} as
    functions of $\rho = m_c/m_b$ for $\mu_b = m_b$.}
    \label{fig:plotsGij_b2ccs}
\end{figure}

In Fig.~\ref{fig:plotsGij_b2ccs} the results for $G_{ij}^{(1)}$
and $G_{ij}^{(2)}$ are shown for $\rho\in[0,0.5]$.
Evaluating the renormalized amplitude for $\mu_b=m_b^{\mathrm{OS}}$ 
we obtain the following numerical values for the decay width
\begin{align}
    \Gamma^{c s c } &= \Gamma_0 
    \left[ 0.86706 
    +3.15768 \frac{\alpha_s}{\pi}
    +37.3426\left(\frac{\alpha_s}{\pi}\right)^2 
    \right] ,
\end{align}

\begin{figure}
    \centering
\includegraphics[width=0.8\textwidth]{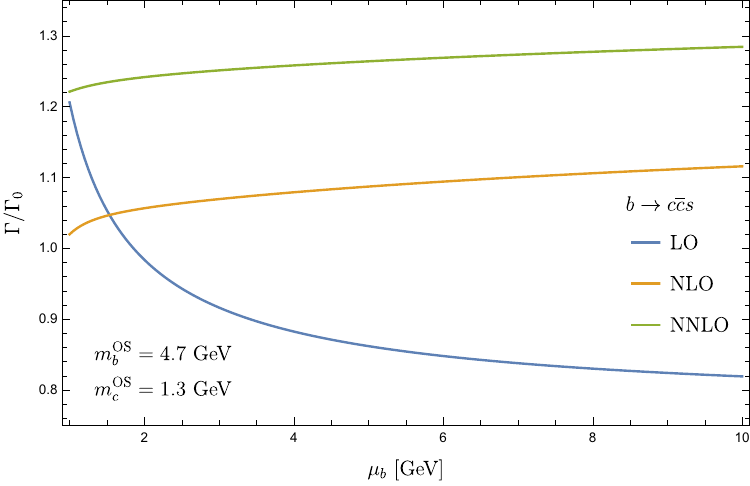}
    \caption{The dependence of the rate for $b \to c \bar c s$ on the renormalization 
    scale $\mu_b \sim m_b$ at LO, NLO and NNLO in the on-shell scheme.}
    \label{fig:mus_dependence_b2ccs}
\end{figure}

In Fig.~\ref{fig:mus_dependence_b2ccs} we show the dependence on the
renormalization scale $\mu_b$. Already at NLO we observe a quite flat
behaviour with a scale variation below 2.5\%. It gets further reduced
to about 1.5\% at NNLO. 
As observed already in~\cite{Voloshin:1994sn,Bagan:1994zd,Krinner:2013cja},
the $O(\alpha_s)$ corrections are rather large, about 25\% of the LO prediction at $\mu_b=m_b$, 
and similarly the $O(\alpha_s^2)$ corrections are 16\% of the NLO prediction.
Note that there is no overlap of the uncertainty band in the considered range of $\mu_b$.
The NLO and NNLO prediction would differ by more than 5 sigma if the theoretical 
uncertainty is entirely based on the scale variation. 
A more conservative approach for the $b \to c \bar c s$ channel would be to 
take as uncertainty of the NNLO prediction half of the $O(\alpha_s^2)$ corrections,
which amounts to about 8\%.

\subsection{The CKM suppressed channel \boldmath $b \to u \bar c s$}

For the LO contribution  we obtain the  same result as in the $b \rightarrow c \bar ud$ decay channel
in Eq.~(\ref{eq::G_b2cud_LO}).
Starting from NLO, the results differ from the  $b \rightarrow c \bar ud$ case. We obtain
\begin{align}
     G_{11}^{(1)} &= \frac{31}{2} - \frac{554}{3}\rho^2 + \frac{554}{3}\rho^6 - \frac{31}{2}\rho^8 + \pi^2\left(-2 + 16\rho^2 + 48\rho^4 - \frac{16}{3}\rho^6 + \frac{2}{3}\rho^8 + 32\rho^4H_0(\rho)\right) \nonumber \\
     &+ \left(-\frac{62}{3} + \frac{640}{3}\rho^2 - \frac{640}{3}\rho^6 + \frac{62}{3}\rho^8\right)H_{-1}(\rho) + \left(-288\rho^4 - 64\rho^6 + 8\rho^8\right)\left(H_0(\rho)\right)^2 
  \nonumber \\
&+
 \left(-96\rho^2 - 120\rho^4 + \frac{992}{3}\rho^6 - \frac{124}{3}\rho^8\right)H_0(\rho)
  + \left(\frac{62}{3} - \frac{640}{3}\rho^2 + \frac{640}{3}\rho^6 - \frac{62}{3}\rho^8\right)H_1(\rho) \nonumber \\ 
  &+
 \left(8 - 64\rho^2 - 576\rho^4 - 64\rho^6 + 8\rho^8\right)\left[H_{0, 1}(\rho) - H_{0,-1}(\rho)\right] ,\\
     G_{22}^{(1)} &= G_{11}^{(1)} ,\\ 
     G_{12}^{(1)} &= -17 + \frac{932}{9}\rho^2 - \frac{932}{9}\rho^6 + 17\rho^8 +
 \left(-\frac{224}{3}\rho^2 + \frac{2048}{3}\rho^4 + \frac{4384}{9}\rho^6 - \frac{344}{9}\rho^8\right)H_0(\rho) \nonumber \\
 &+
\pi^2\left(-\frac{4}{3} + \frac{80}{3}\rho^2 - \frac{256}{3}\rho^3 + 96\rho^4 - \frac{256}{3}\rho^5 - \frac{16}{9}\rho^6 +
   \frac{20}{9}\rho^8 + \frac{128}{3}\rho^4H_0(\rho) \right) \nonumber \\
   &+ 
 \left(-\frac{100}{9} + \frac{1952}{9}\rho^2 - \frac{1952}{9}\rho^6 + \frac{100}{9}\rho^8 +
   \left(-\frac{32}{3} - \frac{128}{3}\rho^2 - \frac{1024}{3}\rho^3 - 576\rho^4 \right.\right. \nonumber \\
   & \left. \left. - \frac{1024}{3}\rho^5 -
     \frac{128}{3}\rho^6 - \frac{32}{3}\rho^8\right)H_0(\rho)\right)H_{-1}(\rho) +
 \left(-192\rho^4 - 64\rho^6 + 16\rho^8\right)\left(H_0(\rho)\right)^2 \nonumber \\
     &+ 
 \left( \frac{100}{9} - \frac{1952}{9}\rho^2 + \frac{1952}{9}\rho^6 - \frac{100}{9}\rho^8 + 
   \left(\frac{32}{3} + \frac{128}{3}\rho^2 - \frac{1024}{3}\rho^3 + 576\rho^4 \right.  \right. \nonumber \\
   & \left. \left. - \frac{1024}{3}\rho^5 + 
     \frac{128}{3}\rho^6 + \frac{32}{3}\rho^8\right)H_0(\rho)\right)H_1(\rho) \nonumber \\
     &+ 
 \left( \frac{16}{3} + \frac{448}{3}\rho^2 + \frac{1024}{3}\rho^3 + 1152\rho^4 + \frac{1024}{3}\rho^5 + \frac{448}{3}\rho^6 + 
   \frac{16}{3}\rho^8\right)H_{0, -1}(\rho) \nonumber \\
   &+ \left(-\frac{16}{3} - \frac{448}{3}\rho^2 + \frac{1024}{3}\rho^3 - 1152\rho^4 + 
   \frac{1024}{3}\rho^5 - \frac{448}{3}\rho^6 - \frac{16}{3}\rho^8\right)H_{0, 1}(\rho).
\end{align}
One observes, that the coefficients $G_{22}^{(1)}$ and $G_{11}^{(1)}$ are the same in this decay channel. 
The NNLO results expanded around $\rho=0$ read
\begin{align}
    G_{11}^{(2)} &= 13.4947  -24.6740 \rho 
+ \left( -552.675 - 591.011 l_\rho - 8.00000l_\rho^2  \right) \rho^2\nonumber \\
&+\left(3314.34 + 210.551 l_\rho \right) \rho^3
+ \left( -1000.88 + 5447.49 l_\rho - 505.906 l_\rho^2 - 64.0000 l_\rho^3 \right)\rho^4\nonumber \\
&
+ \left( 387.348  + 328.986 l_\rho  \right) \rho^5
+ \left( -2248.02  + 240.114 l_\rho + 1066.01 l_\rho^2 - 264.296 l_\rho^3 \right)\rho^6 \nonumber \\
&
    + \left(-159.907 + 306.177l_\rho \right) \rho^7
+  \left(306.693 - 508.164 l_\rho + 133.563 l_\rho^2 - 66.3703 l_\rho^3 \right) \rho^8
,\\
    G_{22}^{(2)} &= G_{11}^{(2)}, \\
    G_{12}^{(2)} &= -72.8419 -16.4493 \rho
    + \left(3424.97 +1112.31 l_\rho + 5.33333 l_\rho^2 +\right) \rho^2\nonumber \\
&
    +\left(7782.51 +   5404.15l_\rho  \right) \rho^3
    +\left( -72.4205 + 15986.63 l_\rho + 727.426 l_\rho^2 - 42.6666 l_\rho^3\right)\rho^4 \nonumber \\
&
    + \left(-2108.79 + 9301.11 l_\rho \right) \rho^5
    + \left(-8020.41+ 2479.56 l_\rho + 1176.61l_\rho^2 - 311.703 l_\rho^3\right)\rho^6 \nonumber \\
&
    + \left( 629.490 + 1924.62 l_\rho\right) \rho^7
    + \left(-1269.71 - 34.1428 l_\rho - 26.6329 l_\rho^2 + 81.3497l_\rho^3\right) \rho^8. 
\end{align}
The limit for $\rho=0$ coincides 
with the result obtained from $b\to c\bar u d$ and $b \to c \bar c s$ calculations. 

\begin{figure}[t]
    \centering    
    \includegraphics[width=0.8\textwidth]{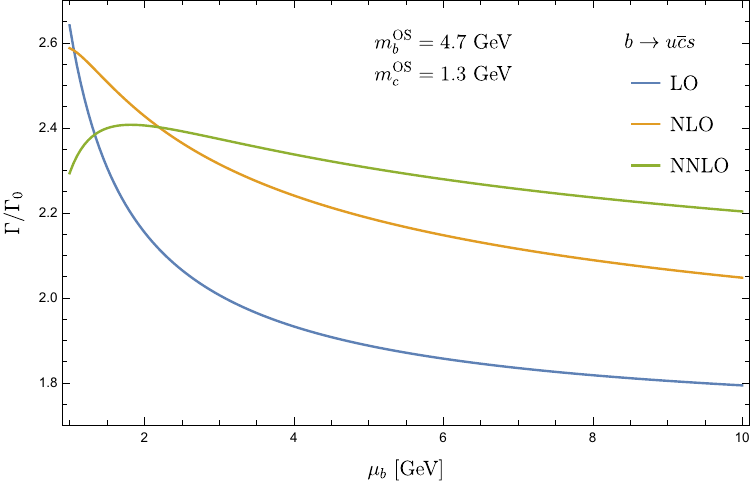}
    \caption{The dependence of the rate for $b \to u \bar c s$ on the renormalization 
    scale $\mu_b \sim m_b^{\mathrm{OS}}$ at LO, NLO and NNLO in the on-shell scheme.}
    \label{fig:mus_dependence_b2ucs}
\end{figure}

At the central scale we obtain the following expansion of the decay rate
\begin{align}
    \Gamma^{uc s} &= \Gamma_0 
    \left[ 1.89907 
    +4.39458 \frac{\alpha_s}{\pi}
    +23.7335\left(\frac{\alpha_s}{\pi}\right)^2 
    \right]\,.
\end{align}
The dependence on $\mu_b$ in Fig.~\ref{fig:mus_dependence_b2ucs} shows
a reduction from 7.3\% to 4.0\% when going from NLO to NNLO.

\subsection{Combined decay channels}

\begin{figure}[t]
    \centering
    \includegraphics[width=0.8\textwidth]{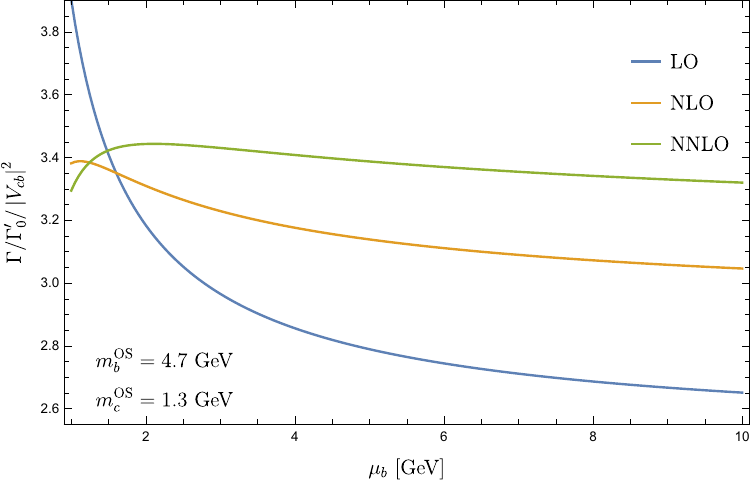}
    \caption{All contributing decay channels combined. }
    \label{fig:mus_dependence_combined}
\end{figure}

The total decay width at partonic level is obtained from the incoherent sum of all individual channels discussed before. 
In particular, we include the contributions from
$b\to c\bar{u}d$,
$b\to c\bar{u}s$,
$b\to c\bar{c}d$,
$b\to c\bar{c}s$,
$b\to u\bar{c}d$,
$b\to u\bar{c}s$,
$b\to u\bar{u}d$ and
$b\to u\bar{u}s$.
The width is in a first approximation given by the sum of the two channel $b \to c \bar u d$ and $b \to c \bar c s$ 
since $|V_{ud}| \simeq |V_{cs}| \simeq 1$. Other channels are CKM suppressed and lead to additional small corrections.
We obtain for $\mu_b=m_b^{\mathrm{OS}}$ the numerical values
\begin{align}
    \Gamma\Big\vert_{\mathrm{all\, nonlep, \, partonic}}
    &= \Gamma^\prime_0 
    \left[ 0.00490995 
    +0.00869123 \frac{\alpha_s}{\pi}
    +0.0898854\left(\frac{\alpha_s}{\pi}\right)^2 
    \right] , 
    \notag \\ &
    =\Gamma_0' |V_{cb}|^2 
    \left[ 
    2.80609 + 4.96713 \frac{\alpha_s}{\pi} + 51.3704 \left(\frac{\alpha_s}{\pi} \right)^2
    \right],
    \label{eqn:GammaAllChannels}
\end{align}
where we define $\Gamma^\prime_0 = G_F^2(m_b^{\mathrm{OS}})^5/(192\pi^3)$ as the normalization of the decay width.
We use the following numerical input for the CKM matrix elements \cite{PDG}:
\begin{align}
    |V_{ud}|&=0.97435 \pm 0.00016,&
    |V_{us}|&=0.22501 \pm 0.00068,&
    |V_{ub}|&=(3.732^{+0.090}_{-0.085})\times 10^{-3}, \notag \\
    |V_{cd}|&=0.22487 \pm 0.00068,&
    |V_{cs}|&=0.97349\pm0.00016 ,&
    |V_{cb}|&=(4.183^{+0.079}_{-0.069}) \times 10^{-2}.
\end{align}
The decay $b \to c \bar u d$ gives about 59\% of the total sum in Eq.~\eqref{eqn:GammaAllChannels},
while $b \to c \bar c s$ contribute about 36\%. The remaining 5\% is given by all other CKM suppressed 
modes.
We observe that the NNLO corrections come with the same sign as the NLO corrections in the relevant region of the renormalization scale $\mu_b$. Using \texttt{RunDec}, we obtain $\alpha_s(\mu_b=4.7\, \mathrm{GeV}) = 0.2166$ and find that the term of $O(\alpha_s^2)$
is roughly 50-60\% of the one at $O(\alpha_s)$ for $\mu_b=m_b^{\mathrm{OS}}$.
The scale uncertainty reduces from a relative 3.5\% at NLO to 1.7\% at NNLO. 
From Fig.~\ref{fig:mus_dependence_combined} we observe that the NNLO curve is flatter than the LO and NLO ones,
however also in the sum of all channels, the predictions at NLO and NNLO differs by about 2$\sim$sigma,
once the theoretical uncertainties are evaluated from the scale variation between $m_b/2 < \mu_b<2m_b$.

\section{Conclusions}
\label{sec:conclusions}

In this paper we provide an important contribution to the hadronic $B$ meson decay rate. We compute NNLO corrections to all relevant partonic channels taking into account finite bottom and charm quark masses. In the effective theory we take into account the current-current operators together with the relevant evanescent operators such that it is possible to use anti-commuting $\gamma_5$ for our calculation. For the computation of the Feynman integrals we use the ``expand and match'' method. It uses the differential equations for the master integrals in order to construct analytic expansions around properly chosen values for $m_c/m_b$ with high-precision numerical coefficients. This leads to compact expressions which are can be evaluated numerically in a straightforward way. 

We perform a preliminary numerical study of the impact of the NNLO corrections 
in the various channels considered using pole scheme for the quark masses. Overall we find that the theoretical uncertainties 
stemming from the scale variation is reduced by more than a factor of three for $b \to c\bar{u}d$.
The reduction for the channel $b \to c\bar{c}s$ is about a factor of two, however we notice
that the NLO and NNLO predictions do not overlap withing the assigned uncertainties,
due to large corrections arising at order $\alpha_s$ and $\alpha_s^2$.

Our analytic results for all decay modes are provided in electronic form and can be retrieved from the
repository~\cite{progdata,egner_2024_11639756}. An update of the lifetime prediction of $B$ meson is ongoing
where we include the novel NNLO corrections presented in this paper and combine
them with updated prediction for the power suppressed terms and quark masses~\cite{in_prep}.
Our results can also be easily applied to decays of $D$ mesons.


\section*{Acknowledgements}  

We thank Mikolaj Misiak for providing us with the matching coefficients 
for $C_1$ and $C_2$ with explicit dependence on $n_f$.
We also thank A.\ Lenz, M.L.\ Piscopo and A.\ Rusov for discussions
and useful comments.
This research was supported by the Deutsche Forschungsgemeinschaft
(DFG, German Research Foundation) under grant 396021762 --- TRR 257
``Particle Physics Phenomenology after the Higgs Discovery'' and has
received funding from the European Research Council (ERC) under the
European Union’s Horizon 2020 research and innovation programme grant
agreement 101019620 (ERC Advanced Grant TOPUP).  The work of M.F. was
supported by the European Union’s Horizon 2020 research and innovation
program under the Marie Sk\l{}odowska-Curie grant agreement
No.~101065445 - PHOBIDE.  


\begin{appendix}


\section{\label{sec::OpMix}Operator mixing}
\subsection{Calculation of renormalization constants}

For our calculation we need the renormalization constants for the mixing of
the effective operators $O_1$ and $O_2$ in the historical basis
up to two loops.  The results can be found in
Refs.~\cite{Buras:1991jm,Buras:1992tc,Ciuchini:1993vr,Chetyrkin:1997gb}, however, not in the form suitable for our
calculation, which is performed keeping the full dependence on $N_C$. For this
reason we decided to repeat the calculation in our setup, which is also a good
check on the correct implementation of the effective operators.  Note that
higher-order results for the renormalization constants and anomalous dimensions
are available in the CMM
basis~\cite{Chetyrkin:1996vx,Gambino:2003zm,Gorbahn:2004my,Gorbahn:2005sa,Czakon:2006ss}. 
Furthermore, transformation rules allow to convert the results from
the historical to the CMM basis and vice versa~\cite{Chetyrkin:1997gb,Gorbahn:2003zz,Gorbahn:2004my}.

In the following we briefly describe the calculation of the renormalization constants
for the physical and evanescent operators appearing in our calculation. 
We consider the matrix element $A_{\mathrm{eff}} = \sum_i C_i \bra{cd} O_i \ket{bu}$ of the effective operators 
with four external quarks (e.g.\ $b u\to c d$).
At higher order in $\alpha_s$, $A_{\mathrm{eff}}$ contains ultraviolet (UV) poles after renormalization 
of the masses, the strong coupling constant and the quark wave functions.
They must be subtracted via the renormalization of the Wilson coefficients,
\begin{align}
C_{i,B}  = Z_{ji} C_j (\mu_b),
\end{align}
where $C_{i,B}$ and $C_i (\mu_b)$ denote the bare and renormalized Wilson coefficients, respectively.
Schematically, the matrix element of the operators is given by
\begin{align}
  A_{\mathrm{eff}} = C_j ( \mu_b) Z_{ji} \langle Z(O_i) \rangle_R,
  \label{eq::Aeff}
\end{align}
where $\langle Z(O_i) \rangle_R$ describes
the expectation value of the operator $O_i$ including
wave function, quark mass and coupling constant renormalization.
The renormalization constants $Z_{ij}$ can be obtained order by order in $\alpha_s$ by 
requiring that Eq.~\eqref{eq::Aeff} is free of UV poles.

The renormalization constants $Z_{ij}$ have the perturbative expansion
\begin{align}
Z_{ij} = \delta_{ij} + \sum_{k=1}^\infty \left( \frac{\alpha_s}{4 \pi} \right)^k Z_{ij}^{(k)},
\end{align}
where 
\begin{align}
Z_{ij}^{(k)} = \sum_{l=0}^k \frac{Z_{ij}^{(k,l)}}{\epsilon^l}.
\end{align}
We use the $\overline{\mathrm{MS}}$ renormalization scheme, which implies
$l>0$. This is, however, not the case if $i$ is the coefficient of an evanescent
operator, while $j$ corresponds to a physical one. In these cases, the
renormalization constants include $\epsilon$-finite terms that ensure that the
matrix elements of evanescent operators vanish in four dimensions
(see Ref.~\cite{Herrlich:1994kh,Misiak:1999yg}).

\begin{figure}
  \centering
  \begin{subfigure}[a]{0.26\textwidth}
    \centering
    \includegraphics[width=0.75\textwidth]{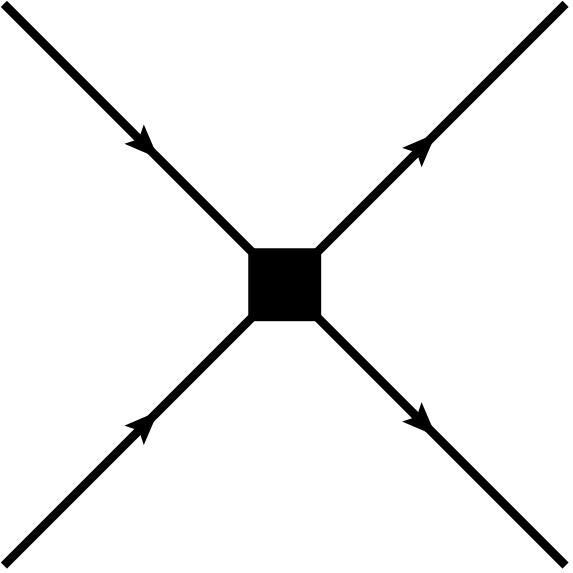}
  \end{subfigure}
  \begin{subfigure}[a]{0.26\textwidth}
    \centering
    \includegraphics[width=0.75\textwidth]{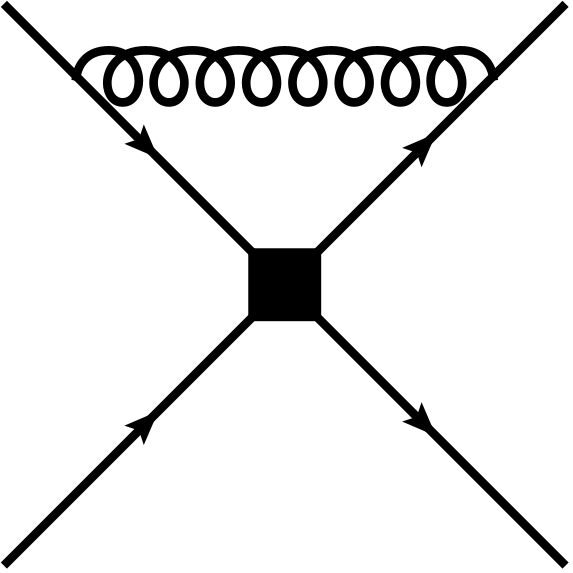}
  \end{subfigure}
  \begin{subfigure}[a]{0.26\textwidth}
    \centering
    \includegraphics[width=0.75\textwidth]{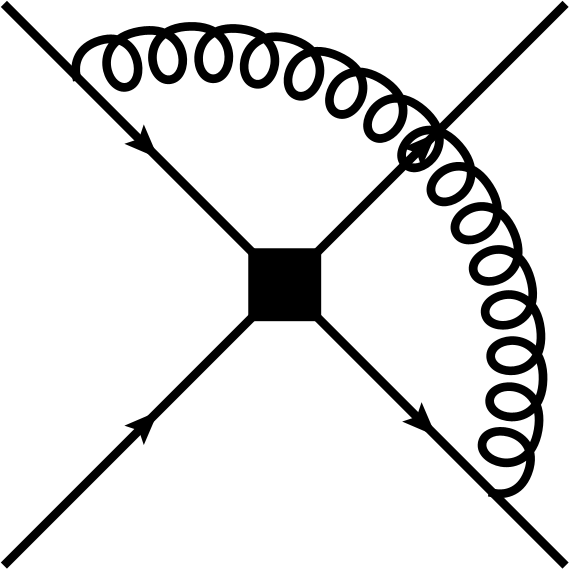}
  \end{subfigure}\\[10pt]
  \begin{subfigure}[a]{0.26\textwidth}
    \centering
    \includegraphics[width=0.75\textwidth]{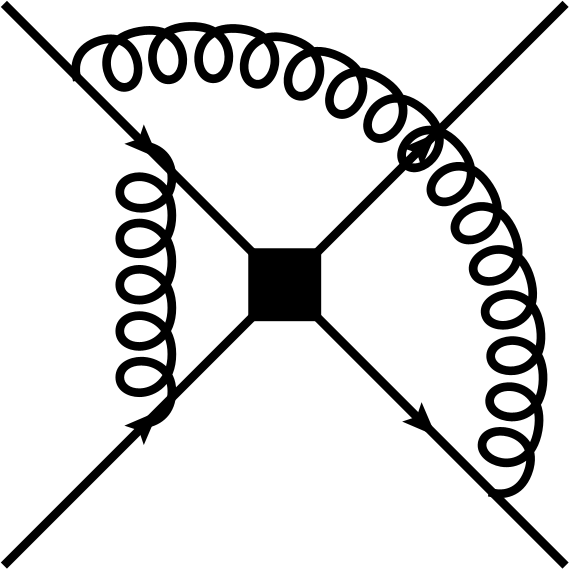}
  \end{subfigure}
  \begin{subfigure}[a]{0.26\textwidth}
    \centering
    \includegraphics[width=0.75\textwidth]{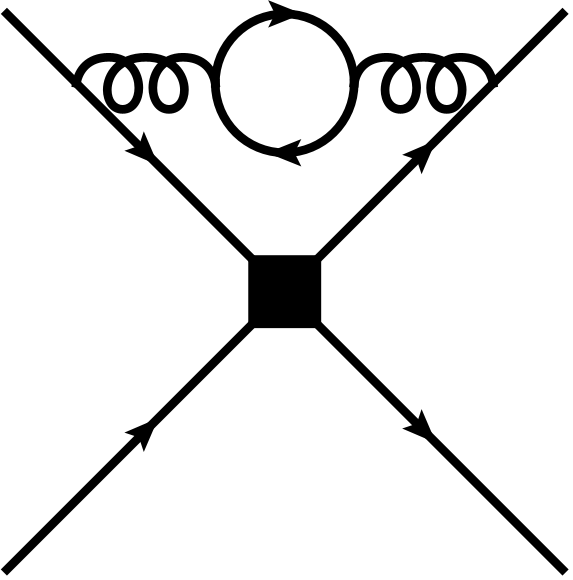}
  \end{subfigure}
  \begin{subfigure}[a]{0.26\textwidth}
    \centering
    \includegraphics[width=0.75\textwidth]{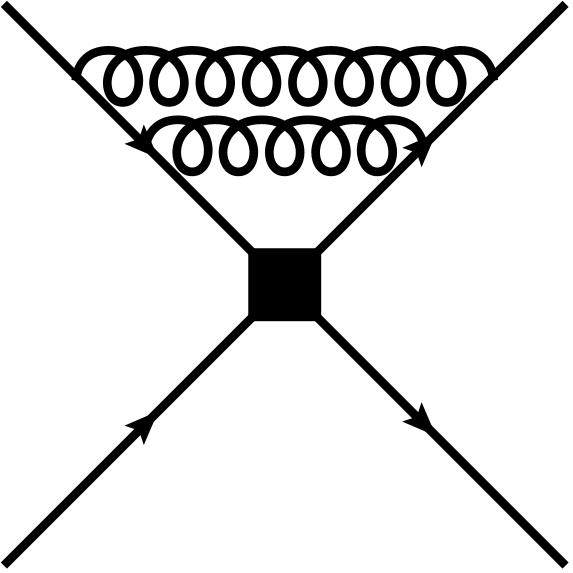}
  \end{subfigure}
  \caption{Sample Feynman diagrams with four external quark lines and the insertion
    of one of the effective operators at tree-level, one- and two-loop order.
    All the quarks are massive and the external momenta are set
    to zero. The square denotes the insertion of an effective four fermion
    operator.
    }
  \label{fig:4FermiOperator}
\end{figure}

For the calculation of the renormalization constants we consider operator
matrix elements for $\{O_1,O_2,E_1^{(1)},E_2^{(1)},E_1^{(2)},E_2^{(2)}\}$
with four external quark lines up two-loop order, see
Fig.~\ref{fig:4FermiOperator} for some sample diagrams.  Since these Feynman
diagrams are logarithmically divergent and since we are only interested in the
UV divergence, we are allowed to choose a convenient kinematic limit
for their computation. In order to avoid infrared divergences, we assign to all
quarks the same mass, keep the gluons and ghosts massless and set the
external momenta to zero.  This leads to one-scale vacuum integrals which are
conveniently computed with the help of {\tt MATAD}~\cite{Steinhauser:2000ry}.

After the calculation of the loop integrals, the result contains terms with up
to nine $\gamma$ matrices and different colour structures. We use the usual
commutation relations and bring the products of $\gamma$ matrices in a
form which allows us to identify the contributions from the physical and
evanescent operators in Eqs.~(\ref{eqn:O1O2definition})
and~(\ref{eqn:evaoperators}), respectively.  For our calculation we need in
addition higher order evanescent operators given by
\begin{align}
  E_5 =& \left( \bar{q}_1^i
         \gamma^{\mu_1 \dots \mu_7}
         P_L b^j \right) \left( \bar{q}_2^i
         \gamma_{\mu_1 \dots \mu_7}
         P_L q_3^j \right)-(4096-7680\epsilon) O_1, \nonumber \\ 
  E_6 =& \left( \bar{q}_1^i
         \gamma^{\mu_1 \dots \mu_7}
         P_L b^i \right) \left( \bar{q}_2^j
         \gamma_{\mu_1 \dots \mu_7}
         P_L q_3^j \right) -(4096-7680\epsilon) O_2, \nonumber \\ 
  E_7 =& \left( \bar{q}_1^i
         \gamma^{\mu_1 \dots \mu_9}
         P_L b^j \right) \left( \bar{q}_2^i
         \gamma_{\mu_1 \dots \mu_9}
         P_L q_3^j \right)
       -(65536-176128\epsilon) O_1, \nonumber \\
  E_8 =& \left( \bar{q}_1^i
         \gamma^{\mu_1 \dots \mu_9}
         P_L b^i \right) \left( \bar{q}_2^j
         \gamma_{\mu_1 \dots \mu_9}
         P_L q_3^j \right) 
       -(65536-176128\epsilon) O_2.
\end{align}
The $O(\epsilon)$ terms can be obtained following Ref.~\cite{Buras:1998raa};
in the case of $E_5$ and $E_6$ they can also be found in Ref.~\cite{Buras:2006gb}.

Once the bare result is expressed as a linear combination of (bare) operator
matrix elements, we perform the parameter renormalization (wave function, quark
mass and strong coupling constant) in the $\overline{\rm MS}$ scheme and
introduce the $Z_{ij}$ according to Eq.~(\ref{eq::Aeff}), where the unknown
coefficients are obtained from the requirement that the renormalized operator
matrix elements are finite.

Our results for the one-loop renormalization constants for the following set of physical and evanescent operators
\begin{align}
    \vec Q^T &= (O_1, O_2), &
    \vec E^T &= (E_1^{(1)}, E_2^{(1)},E_1^{(2)}, E_2^{(2)} ) ,
\end{align}
are given by
\begin{align}
  Z^{(1,1)} &= \left( \begin{array}{c c c c c c}
                        -1 & 3 & \frac{7}{12} & \frac{1}{4} & 0 & 0 \\[5pt]
                        3 & -1 & \frac{1}{2} & - \frac{1}{6} & 0 & 0 \\[5pt]
                        0 & 0 & - \frac{59}{3} & -5 & \frac{7}{12} & \frac{1}{4} \\[5pt]
                        0 & 0 & - 13 & \frac{13}{3} & \frac{1}{2} & - \frac{1}{6} \\ [5pt]
                        0 & 0 & - \frac{1888}{3} & 96 & \frac{41}{3} & - 9 \\[5pt]
                        0 & 0 & - 288 & \frac{1568}{3} & 3 & - \frac{67}{3}
                      \end{array} \right), \nonumber \\[5pt]
  Z^{(1,0)} &= \left( \begin{array}{c c c c c c}
                        0 & 0 & 0 & 0 & 0 & 0 \\[5pt]
                        0 & 0 & 0 & 0 & 0 & 0 \\[5pt]
                        16 & -48 & 0 & 0 & 0 & 0 \\[5pt]
                        -24 & -56 & 0 & 0 & 0 & 0 \\[5pt]
                        512 & -1536 & 0 & 0 & 0 & 0 \\[5pt]
                        -768 & -1792 & 0 & 0 & 0 & 0 
                      \end{array} \right)\,,
                                                   \label{eq:ZNLOTrad}
\end{align}
and at two loops we have
{\small
\begin{align}
  Z^{(2,2)} &= 
  \left(
\begin{array}{cccccc}
 \frac{21}{2} & -\frac{39}{2} & -\frac{91}{9} & -\frac{8}{3} & \frac{67}{288} & \frac{5}{96} \\[5pt]
 -\frac{39}{2} & \frac{21}{2} & -\frac{143}{24} & -\frac{17}{72} & \frac{5}{48} & \frac{11}{144} \\[5pt]
 0 & 0 & \frac{229}{2} & \frac{955}{6} & -\frac{35}{6} & -\frac{53}{6} \\[5pt]
 0 & 0 & \frac{227}{6} & -\frac{3}{2} & -\frac{55}{24} & -\frac{35}{24} \\[5pt]
 0 & 0 & \frac{13360}{9} & \frac{5200}{3} & -\frac{2273}{6} & \frac{907}{6} \\[5pt]
 0 & 0 & \frac{1424}{3} & -\frac{63248}{9} & -\frac{793}{6} & \frac{1675}{6} 
\end{array}
\right)
  \notag \\[10pt] &
  +n_f \left(
\begin{array}{cccccc}
 -\frac{1}{3} & 1 & \frac{7}{36} & \frac{1}{12} & 0 & 0 \\[5pt]
 1 & -\frac{1}{3} & \frac{1}{6} & -\frac{1}{18} & 0 & 0 \\[5pt]
 0 & 0 & -\frac{59}{9} & -\frac{5}{3} & \frac{7}{36} & \frac{1}{12} \\[5pt]
 0 & 0 & -\frac{13}{3} & \frac{13}{9} & \frac{1}{6} & -\frac{1}{18} \\[5pt]
 0 & 0 & -\frac{1888}{9} & 32 & \frac{41}{9} & -3 \\[5pt]
 0 & 0 & -96 & \frac{1568}{9} & 1 & -\frac{67}{9} 
\end{array}
\right),
\notag \\[10pt]
  Z^{(2,1)} &=
 \left(
\begin{array}{cccccc}
 -\frac{23}{24} & -\frac{161}{8} & \frac{323}{36} & \frac{7}{12} & -\frac{1}{36} & -\frac{5}{48} \\[5pt]
 \frac{55}{8} & -\frac{239}{24} & \frac{51}{8} & \frac{115}{72} & -\frac{35}{384} & -\frac{77}{1152} \\[5pt]
 -212 & 252 & -\frac{1279}{24} & -\frac{697}{8} & \frac{145}{24} & \frac{49}{24} \\[5pt]
 96 & 256 & -\frac{1963}{24} & \frac{2753}{24} & \frac{89}{12} & -\frac{1}{12} \\[5pt]
 -3968 & 11904 & \frac{32488}{9} & 1704 & \frac{9257}{24} & -\frac{2817}{8} \\[5pt]
 4992 & 16768 & -\frac{12748}{3} & \frac{101644}{9} & \frac{6985}{24} & \frac{2117}{24} 
\end{array}
\right)
  \nonumber \\[10pt]
  &+ n_f
  \left(
\begin{array}{cccccc}
 -\frac{1}{18} & \frac{1}{6} & -\frac{7}{216} & -\frac{1}{72} & 0 & 0 \\[5pt]
 \frac{1}{6} & -\frac{1}{18} & -\frac{1}{36} & \frac{1}{108} & 0 & 0 \\[5pt]
 \frac{16}{3} & -16 & \frac{353}{54} & -\frac{1}{18} & -\frac{7}{216} & -\frac{1}{72} \\[5pt]
 -8 & -\frac{56}{3} & \frac{67}{18} & -\frac{259}{54} & -\frac{1}{36} & \frac{1}{108} \\[5pt]
 \frac{512}{3} & -512 & \frac{4280}{27} & -\frac{40}{3} & \frac{427}{54} & -\frac{3}{2} \\[5pt]
 -256 & -\frac{1792}{3} & 80 & -\frac{3280}{27} & \frac{7}{2} & -\frac{383}{54} 
\end{array}
\right), \notag \\[10pt]
  Z^{(2,0)} &= 
  \left(
\begin{array}{cccccc}
 0 & 0 & 0 & 0 & 0 & 0 \\
 0 & 0 & 0 & 0 & 0 & 0 \\
 \frac{202796}{115}-\frac{50488 n_f}{1035} & \frac{1037132}{115}-\frac{25544 n_f}{345} & 0 & 0 & 0 & 0 \\[5pt]
 -\frac{15148 n_f}{345}-\frac{39856}{115} & \frac{46028 n_f}{1035}+\frac{39904}{23} & 0 & 0 & 0 & 0 \\[5pt]
 \frac{4098848}{1035}-\frac{1561184 n_f}{1035} & \frac{1154912 n_f}{345}-\frac{1011808}{15} & 0 & 0 & 0 & 0 \\[5pt]
 \frac{56192 n_f}{69}-\frac{34695184}{345} & \frac{6995584 n_f}{1035}-\frac{282282736}{1035} & 0 & 0 & 0 & 0 
\end{array}
\right).
\label{eq:ZNNLOTrad}
\end{align}}
\\
We have performed the calculation for general SU($N_c$) gauge groups; the corresponding analytic expressions can be found in the supplementary material to this paper~\cite{progdata,egner_2024_11639756}. For 
simplicity we present the results for $N_c=3$ and only keep the the number
of active flavours $n_f$ in the analytic expressions. In the numerical analysis we use $n_f=5$.

At first sight it looks strange that with our approach also the finite terms
in $Z^{(1,0)}$ and $Z^{(2,0)}$ can be determined. Note, however, that
they originate from divergent part of loop integrals multiplied by factor
of $\epsilon$ from the Dirac algebra in the numerator of the Feynman
diagrams.
Our calculation has been performed for general QCD gauge parameter which drops
out in the final results for the renormalization constants.  Furthermore,
after converting the renormalization constants to the anomalous dimension
matrix, we agree with the results in Eq.\ (57) of Ref.~\cite{Chetyrkin:1997gb}.

Our setup is validated also by reproducing the well known results in the CMM basis.
For the calculation of the $Z_{ij}$ in the CMM basis, we use the physical
operators from Eq.~(\ref{eqn:O1O2CMMdefinition}). The evanescent operators are
given by
\begin{align}
  E_1' &= \left( \bar{q}_1 \gamma^{\mu_1 \mu_2 \mu_3} T^a P_L b \right) 
  \left( \bar{q}_2 \gamma_{\mu_1 \mu_2 \mu_3} T^a P_L q_3 \right) - 16 O_1',\nonumber \\
  E_2' &= \left( \bar{q}_1 \gamma^{\mu_1 \mu_2 \mu_3}  P_L b \right) 
  \left( \bar{q}_2 \gamma_{\mu_1 \mu_2 \mu_3} P_L q_3 \right) - 16 O_2' ,\nonumber \\
  E_3' &= \left( \bar{q}_1 \gamma^{\mu_1 \dots \mu_5} T^a P_L b \right) 
  \left( \bar{q}_2 \gamma_{\mu_1 \dots \mu_5} T^a P_L q_3 \right) - 256 O_1' -20 E_1'  ,\nonumber \\
  E_4' &= \left( \bar{q}_1 \gamma^{\mu_1 \dots \mu_5} P_L b \right) 
  \left( \bar{q}_2 \gamma_{\mu_1 \dots \mu_5}  P_L q_3 \right)   - 256 O_2' -20 E_2'
        \,,\nonumber\\
  E_5' &= \left( \bar{q}_1 \gamma^{\mu_1 \dots \mu_7} T^a P_L b \right) 
  \left( \bar{q}_2 \gamma_{\mu_1 \dots \mu_7} T^a P_L q_3 \right) -4096 O_1' -336E_1',\nonumber \\
  E_6' &= \left( \bar{q}_1 \gamma^{\mu_1 \dots \mu_7} P_L b \right) 
  \left( \bar{q}_2 \gamma_{\mu_1 \dots \mu_7} P_L q_3 \right) -4096 O_2' -336E_2' ,\nonumber \\
  E_7' &= \left( \bar{q}_1 \gamma^{\mu_1 \dots \mu_9} T^a P_L b \right) 
  \left( \bar{q}_2 \gamma_{\mu_1 \dots\mu_9} T^a P_L q_3 \right) -65536O_1' -5440E_1',\nonumber \\
  E_8' &= \left( \bar{q}_1 \gamma^{\mu_1 \dots \mu_9} P_L b \right) 
  \left( \bar{q}_2 \gamma_{\mu_1 \dots \mu_9} P_L q_3 \right)-65536O_2' -5440E_2' \,.
\end{align}
We were able to reproduce the results from Ref.~\cite{Gorbahn:2004my}.
Furthermore we use the formalism developed in Refs.~\cite{Chetyrkin:1997gb,Gorbahn:2003zz}
and translate the renormalization constants from the CMM to the
historical basis (see also next subsection), which confirms the results given 
in Eqs.~\eqref{eq:ZNLOTrad} and~\eqref{eq:ZNNLOTrad}.

\subsection{\label{sub::basis}Change of basis}

In order to calculate the NNLO anomalous dimension in Eq.~\eqref{eqn:gamma2} in the historical basis
fulfilling the condition~\eqref{eqn:conditionsFierz}
and to compare the renormalization constants with known results 
for the CMM basis in the literature, we have to perform a basis transformation.
In the following we adopt the formalism developed in~\cite{Chetyrkin:1997gb,Gorbahn:2003zz,Buras:2006gb}
to describe the basis change between the CMM basis and the historical basis.
Let us denote by 
\begin{align}
\vec Q^{' T} &= (O_1', O_2'), \nonumber \\
\vec E^{' T} &= (E_1', E_2', E_3',E_4'),  
\end{align}
the physical and evanescent operators in the CMM basis.
For physical operators, the basis change is a simple
linear transformation
\begin{equation}
  \vec Q = \hat R \vec Q',
\end{equation}
where in our case $\hat R$ is a $2 \times 2$ matrix.\footnote{
In general one has to consider a basis change where also  some evanescent operators
are added to the physical ones, i.e.\ 
$
  \vec Q ' = \hat R ( \vec Q + \hat W \vec E),
$
where $\hat W$ is a $2 \times n$ matrix. In our case $\hat W = 0$.}
For the evanescent operators, the transformation rule is 
\begin{equation}
  \vec E = \hat M \Big[ \vec E' + \epsilon \hat U \vec Q '+ \epsilon^2 \hat V \vec Q' \Big],
\end{equation}
where the $n\times 2$ matrices $\hat U$ and $\hat V$ and the $n\times n$
matrix $\hat M$ parametrize the rotation of evanescent operators.
The basis change for the operator set $(\vec Q, \vec E)$ 
is encoded by two $\epsilon$-dependent linear transformations,
\begin{align}
  \hat A &= 
  \begin{pmatrix}
  \hat R & 0 \\
  0 & \hat M 
  \end{pmatrix}, &
  \hat B &=   
  \begin{pmatrix}
  1 & 0 \\
  \epsilon \hat U + \epsilon^2 \hat V & 1 
  \end{pmatrix}, 
\end{align}
so that renormalization constants in the two bases are related by
\begin{equation}
  \hat Z =
  (\hat A \hat B) Z' (\hat B^{-1} \hat A^{-1}).
\end{equation}
The transformation matrices from the CMM basis to the historical basis 
with the evanescent operators defined in Eq.~\eqref{eqn:evaoperators} and $A_2 = -4$ 
are
\begin{align}
  \hat R &= 
  \begin{pmatrix}
    2 & 1/3 \\
    0 & 1
  \end{pmatrix}, &
  \hat M &=
  \begin{pmatrix}
 2 & \frac{1}{3} & 0 & 0 & 0 & 0 \\
 0 & 1 & 0 & 0 & 0 & 0 \\
 40 & \frac{20}{3} & 2 & \frac{1}{3} & 0 & 0 \\
 0 & 20 & 0 & 1 & 0 & 0 \\
 672 & 112 & 0 & 0 & 2 & \frac{1}{3} \\
 0 & 336 & 0 & 0 & 0 & 1 \\
  \end{pmatrix},
  \notag \\[10pt]
  \hat U & =
  \begin{pmatrix}
   4 & 0 \\
   0 & 4 \\
   144 & 0 \\
   0 & 144 \\
   6336 & 0 \\
   0 & 6336 \\
  \end{pmatrix}, &
  \hat V & =
  \begin{pmatrix}
 4 & 0 \\
 0 & 4 \\[5pt]
 \frac{36736}{115} & -\frac{2304}{23} \\[5pt]
 0 & \frac{105856}{115} \\[5pt]
 -1344 & 0 \\
 0 & -1344 \\
  \end{pmatrix}.
\end{align}

By focusing on the various subblocks of the 
renormalization matrices,
\begin{equation}
  \hat Z = \begin{pmatrix}
  \hat Z_{QQ} & \hat Z_{QE} \\
  \hat Z_{EQ} & \hat Z_{EE} 
\end{pmatrix},
\end{equation}
we can give the transformation rules at order $\alpha_s$
\begin{align}
  \hat Z_{QQ}^{(1,1)} &= 
  \hat R \hat Z^{'(1,1)}_{QQ} \hat R^{-1}, 
  &
  \hat Z_{QE}^{(1,1)} &= 
  \hat R \hat Z^{'(1,1)}_{QE} \hat M^{-1}, 
  \notag \\
  \hat Z_{EE}^{(1,1)} &= \hat M \hat Z^{'(1,1)}_{EE} \hat M^{-1}, 
  &
  \hat Z_{QQ}^{(1,0)} &= - \hat R \hat Z_{QE}^{'(1,1)} \hat U \hat R^{-1},
  \notag \\
  \hat Z_{EQ}^{(1,0)} &=\hat  M 
  \Big[
  \hat  Z^{'(1,0)}_{EQ}  
  +\hat  U \hat Z_{QQ}^{'(1,1)} 
  - \hat Z_{EE}^{'(1,1)} \hat U 
  \Big]\hat R^{-1}. 
\end{align}
At order $\alpha_s^2$ we have
\begin{align}
  \hat Z_{QQ}^{(2,2)} &= 
  \hat R \hat Z^{'(2,2)}_{QQ} \hat R^{-1}, 
  \notag \\
  \hat Z_{QE}^{(2,2)} &= 
  \hat R \hat Z^{'(2,2)}_{QE} \hat M^{-1}, 
  \notag \\
  \hat Z_{EE}^{(2,2)} &= 
  \hat M \hat Z^{'(2,2)}_{EE} \hat M^{-1}, 
  \notag \\ 
  \hat Z_{QQ}^{(2,1)} &= 
  \hat R \Big[ \hat Z^{'(2,1)}_{QQ} 
  -\hat Z^{'(2,2)}_{QE} \hat U \Big] \hat R^{-1}, 
  \notag \\ 
  \hat Z_{QE}^{(2,1)} &=
  \hat R \hat Z^{'(2,1)}_{QE} \hat M^{-1},
  \notag \\
  \hat Z_{EE}^{(2,1)} &=
  \hat R \Big[ 
  \hat Z^{'(2,1)}_{EE} 
  -\hat U \hat Z^{'(2,2)}_{QE}
  \Big] \hat M^{-1},
  \notag \\
  \hat Z_{EQ}^{(2,1)} &=
  \hat M \Big[ 
  \hat Z^{'(2,1)}_{EQ} 
  +\hat U \hat Z^{'(2,2)}_{QQ}
  -\hat Z^{'(2,2)}_{EE} \hat U
  \Big] \hat R^{-1},
  \notag \\
  \hat Z_{QQ}^{(2,0)} &=
  \hat R \Big[ 
  -\hat Z^{'(2,1)}_{QE} \hat U
  -\hat Z^{'(2,2)}_{QE} \hat V
  +{ \hat Z_{QE}^{'(1,1)} }\hat V \hat Z_{QQ}^{'(1,1)} 
  \Big] \hat R^{-1} 
  \notag \\
  \hat Z_{EQ}^{(2,0)} &=
  \hat M \Big[ 
    \hat Z^{'(2,0)}_{EQ}
    +\hat U \hat Z^{'(2,1)}_{QQ}
    +\hat V \hat Z^{'(2,2)}_{QQ}
    -\hat Z^{'(2,1)}_{EE} \hat U
    -\hat Z^{'(2,2)}_{EE} \hat V
    -\hat U \hat Z^{'(2,2)}_{QE} \hat U
  \Big] \hat R^{-1}.
\end{align}
After rotating the CMM basis into the historical basis, the element $\hat Z^{(1,0)}_{QQ}$ and $\hat Z^{(2,0)}_{QQ}$ 
are different from zero and therefore they do not corresponds to an $\overline{\mathrm{MS}}$ renormalization scheme.
Such finite contributions must be removed by a suitable change of scheme.
For the renormalization constants this corresponds to the transformation
\begin{equation}
    \hat Z_{\overline{\mathrm{MS}}} =
    \left[
    1
    - \frac{\alpha_s}{4 \pi}
    \hat r_1
    - \left(\frac{\alpha_s}{4 \pi}\right)^2
    (\hat r_2-\hat r_1^2)
    \right]
    \hat Z,
\end{equation}
where for the subblock corresponding to the physical operators we obtain
\begin{align}
    (\hat r_1)_{QQ} &= \hat Z_{QQ}^{(1,0)} =
    \begin{pmatrix}
        -\frac{7}{3} & -1 \\
        -2 & \frac{2}{3}
    \end{pmatrix}, 
    \notag \\[5pt]
    (\hat r_2)_{QQ} &= \hat Z_{QQ}^{(2,0)} =
    \begin{pmatrix}
    -\frac{153257}{2070}-\frac{35 }{54} n_f & -\frac{1763}{138}-\frac{5 }{18} n_f \\[5pt]
    -\frac{6493}{276}-\frac{5 }{9} n_f & -\frac{239239}{4140}+\frac{5 }{27} n_f 
   \end{pmatrix}.
\end{align}
The transformation rule for the Wilson coefficients is
\begin{equation}
    \vec C(\mu_b) = 
    \vec C'(\mu_b) \hat R^{-1}
    \left[
    1
    + \frac{\alpha_s}{4 \pi}
    (\hat r_1)_{QQ}
    + \left(\frac{\alpha_s}{4 \pi}\right)^2
    (\hat r_2)_{QQ}
    \right],
    \label{eqn:basis_change_Wilsoncoefficients}
\end{equation}
while the ADMs transform in the following way:
\begin{align}
    \hat \gamma^{(0)} &= 
    \hat R \hat \gamma^{'(0)} \hat R^{-1},
    \notag \\
    \hat \gamma^{(1)} &= 
    \hat R \hat \gamma^{'(1)} \hat R^{-1}
    -\Big[ \hat Z^{(1,0)}_{QQ} , \hat \gamma^{(0)} \Big]
    -2 \beta_0 \hat Z^{(1,0)},
    \notag \\
    \hat \gamma^{(2)} &= 
    \hat R \hat \gamma^{'(2)} \hat R^{-1}
    -\Big[ (\hat r_2)_{QQ} , \hat \gamma^{(0)} \Big]
    -\Big[ \hat Z^{(1,0)}_{QQ} , \hat \gamma^{(1)} \Big]
    +\Big[ \hat Z^{(1,0)}_{QQ} , \hat \gamma^{(0)} \Big] \hat Z^{(1,0)}_{QQ}
    \notag \\
    & \quad -4\beta_0 (\hat r_2)_{QQ}
    -2 \beta_1 \hat Z^{(1,0)}_{QQ}
    +2 \beta_0 (\hat Z^{(1,0)}_{QQ})^2.
\end{align}
Our expressions for the NNLO anomalous dimension in the historical basis and the evanescent operator definition given in Eq.~\eqref{eqn:evaoperators} 
with $A_2 = -4$ is shown in Eq.~\eqref{eqn:gamma2}
which fulfils the condition~\eqref{eqn:conditionsFierz}.

\end{appendix}

\bibliographystyle{jhep} 
\bibliography{b2cud.bib,noninspires.bib}

\providecommand{\href}[2]{#2}\begingroup\raggedright\begin{thebibliography}{10}

\bibitem{Lenz:2014jha}
A.~Lenz, \emph{{Lifetimes and heavy quark expansion}},
  \href{https://doi.org/10.1142/S0217751X15430058}{\emph{Int. J. Mod. Phys. A}
  {\bfseries 30} (2015) 1543005}
  [\href{https://arxiv.org/abs/1405.3601}{{\ttfamily 1405.3601}}].

\bibitem{Albrecht:2024oyn}
J.~Albrecht, F.~Bernlochner, A.~Lenz and A.~Rusov, \emph{{Lifetimes of
  b-hadrons and mixing of neutral B-mesons: theoretical and experimental
  status}}, \href{https://doi.org/10.1140/epjs/s11734-024-01124-3}{\emph{Eur.
  Phys. J. ST} {\bfseries 233} (2024) 359}
  [\href{https://arxiv.org/abs/2402.04224}{{\ttfamily 2402.04224}}].

\bibitem{Bernlochner:2022ucr}
F.~Bernlochner, M.~Fael, K.~Olschewsky, E.~Persson, R.~van Tonder, K.~K. Vos
  and M.~Welsch, \emph{{First extraction of inclusive V$_{cb}$ from q$^{2}$
  moments}}, \href{https://doi.org/10.1007/JHEP10(2022)068}{\emph{JHEP}
  {\bfseries 10} (2022) 068}
  [\href{https://arxiv.org/abs/2205.10274}{{\ttfamily 2205.10274}}].

\bibitem{Finauri:2023kte}
G.~Finauri and P.~Gambino, \emph{{The q$^{2}$ moments in inclusive semileptonic
  B decays}}, \href{https://doi.org/10.1007/JHEP02(2024)206}{\emph{JHEP}
  {\bfseries 02} (2024) 206}
  [\href{https://arxiv.org/abs/2310.20324}{{\ttfamily 2310.20324}}].

\bibitem{Kirk:2017juj}
M.~Kirk, A.~Lenz and T.~Rauh, \emph{{Dimension-six matrix elements for meson
  mixing and lifetimes from sum rules}},
  \href{https://doi.org/10.1007/JHEP12(2017)068}{\emph{JHEP} {\bfseries 12}
  (2017) 068} [\href{https://arxiv.org/abs/1711.02100}{{\ttfamily
  1711.02100}}].

\bibitem{King:2021jsq}
D.~King, A.~Lenz and T.~Rauh, \emph{{SU(3) breaking effects in B and D meson
  lifetimes}}, \href{https://doi.org/10.1007/JHEP06(2022)134}{\emph{JHEP}
  {\bfseries 06} (2022) 134}
  [\href{https://arxiv.org/abs/2112.03691}{{\ttfamily 2112.03691}}].

\bibitem{Lin:2022fun}
J.~Lin, W.~Detmold and S.~Meinel, \emph{{Lattice Study of Spectator Effects in
  $b$-hadron Decays}}, \href{https://doi.org/10.22323/1.430.0417}{\emph{PoS}
  {\bfseries LATTICE2022} (2023) 417}
  [\href{https://arxiv.org/abs/2212.09275}{{\ttfamily 2212.09275}}].

\bibitem{Black:2023vju}
M.~Black, R.~Harlander, F.~Lange, A.~Rago, A.~Shindler and O.~Witzel,
  \emph{{Using Gradient Flow to Renormalise Matrix Elements for Meson Mixing
  and Lifetimes}}, \href{https://doi.org/10.22323/1.453.0263}{\emph{PoS}
  {\bfseries LATTICE2023} (2024) 263}
  [\href{https://arxiv.org/abs/2310.18059}{{\ttfamily 2310.18059}}].

\bibitem{Pak:2008qt}
A.~Pak and A.~Czarnecki, \emph{{Mass effects in muon and semileptonic b
  ---\ensuremath{>} c decays}},
  \href{https://doi.org/10.1103/PhysRevLett.100.241807}{\emph{Phys. Rev. Lett.}
  {\bfseries 100} (2008) 241807}
  [\href{https://arxiv.org/abs/0803.0960}{{\ttfamily 0803.0960}}].

\bibitem{Pak:2008cp}
A.~Pak and A.~Czarnecki, \emph{{Heavy-to-heavy quark decays at NNLO}},
  \href{https://doi.org/10.1103/PhysRevD.78.114015}{\emph{Phys. Rev. D}
  {\bfseries 78} (2008) 114015}
  [\href{https://arxiv.org/abs/0808.3509}{{\ttfamily 0808.3509}}].

\bibitem{Dowling:2008mc}
M.~Dowling, J.~H. Piclum and A.~Czarnecki, \emph{{Semileptonic decays in the
  limit of a heavy daughter quark}},
  \href{https://doi.org/10.1103/PhysRevD.78.074024}{\emph{Phys. Rev. D}
  {\bfseries 78} (2008) 074024}
  [\href{https://arxiv.org/abs/0810.0543}{{\ttfamily 0810.0543}}].

\bibitem{Melnikov:2008qs}
K.~Melnikov, \emph{{O(alpha(s)**2) corrections to semileptonic decay b
  ---\ensuremath{>} cl anti-nu(l)}},
  \href{https://doi.org/10.1016/j.physletb.2008.07.089}{\emph{Phys. Lett. B}
  {\bfseries 666} (2008) 336}
  [\href{https://arxiv.org/abs/0803.0951}{{\ttfamily 0803.0951}}].

\bibitem{Fael:2020tow}
M.~Fael, K.~Sch\"onwald and M.~Steinhauser, \emph{{Third order corrections to
  the semileptonic b\textrightarrow{}c and the muon decays}},
  \href{https://doi.org/10.1103/PhysRevD.104.016003}{\emph{Phys. Rev. D}
  {\bfseries 104} (2021) 016003}
  [\href{https://arxiv.org/abs/2011.13654}{{\ttfamily 2011.13654}}].

\bibitem{Fael:2023tcv}
M.~Fael and J.~Usovitsch, \emph{{Third order correction to semileptonic $b\to
  u$ decay: Fermionic contributions}},
  \href{https://doi.org/10.1103/PhysRevD.108.114026}{\emph{Phys. Rev. D}
  {\bfseries 108} (2023) 114026}
  [\href{https://arxiv.org/abs/2310.03685}{{\ttfamily 2310.03685}}].

\bibitem{Fael:2022frj}
M.~Fael, K.~Sch\"onwald and M.~Steinhauser, \emph{{A first glance to the
  kinematic moments of B \textrightarrow{}
  X$_{c}$\ensuremath{\ell}\ensuremath{\nu} at third order}},
  \href{https://doi.org/10.1007/JHEP08(2022)039}{\emph{JHEP} {\bfseries 08}
  (2022) 039} [\href{https://arxiv.org/abs/2205.03410}{{\ttfamily
  2205.03410}}].

\bibitem{Buras:1989xd}
A.~J. Buras and P.~H. Weisz, \emph{{QCD Nonleading Corrections to Weak Decays
  in Dimensional Regularization and 't Hooft-Veltman Schemes}},
  \href{https://doi.org/10.1016/0550-3213(90)90223-Z}{\emph{Nucl. Phys. B}
  {\bfseries 333} (1990) 66}.

\bibitem{Buchalla:1995vs}
G.~Buchalla, A.~J. Buras and M.~E. Lautenbacher, \emph{{Weak decays beyond
  leading logarithms}},
  \href{https://doi.org/10.1103/RevModPhys.68.1125}{\emph{Rev. Mod. Phys.}
  {\bfseries 68} (1996) 1125}
  [\href{https://arxiv.org/abs/hep-ph/9512380}{{\ttfamily hep-ph/9512380}}].

\bibitem{Buras:2011we}
A.~J. Buras, \emph{{Climbing NLO and NNLO Summits of Weak Decays: 1988-2023}},
  \href{https://doi.org/10.1016/j.physrep.2023.07.002}{\emph{Phys. Rept.}
  {\bfseries 1025} (2023) } [\href{https://arxiv.org/abs/1102.5650}{{\ttfamily
  1102.5650}}].

\bibitem{Altarelli:1991dx}
G.~Altarelli and S.~Petrarca, \emph{{Inclusive beauty decays and the spectator
  model}}, \href{https://doi.org/10.1016/0370-2693(91)90332-K}{\emph{Phys.
  Lett. B} {\bfseries 261} (1991) 303}.

\bibitem{Buchalla:1992gc}
G.~Buchalla, \emph{{O (alpha-s) QCD corrections to charm quark decay in
  dimensional regularization with nonanticommuting gamma-5}},
  \href{https://doi.org/10.1016/0550-3213(93)90081-Y}{\emph{Nucl. Phys. B}
  {\bfseries 391} (1993) 501}.

\bibitem{Bagan:1994zd}
E.~Bagan, P.~Ball, V.~M. Braun and P.~Gosdzinsky, \emph{{Charm quark mass
  dependence of QCD corrections to nonleptonic inclusive B decays}},
  \href{https://doi.org/10.1016/0550-3213(94)90591-6}{\emph{Nucl. Phys. B}
  {\bfseries 432} (1994) 3}
  [\href{https://arxiv.org/abs/hep-ph/9408306}{{\ttfamily hep-ph/9408306}}].

\bibitem{Bagan:1995yf}
E.~Bagan, P.~Ball, B.~Fiol and P.~Gosdzinsky, \emph{{Next-to-leading order
  radiative corrections to the decay b ---\ensuremath{>} c c s}},
  \href{https://doi.org/10.1016/0370-2693(95)00437-P}{\emph{Phys. Lett. B}
  {\bfseries 351} (1995) 546}
  [\href{https://arxiv.org/abs/hep-ph/9502338}{{\ttfamily hep-ph/9502338}}].

\bibitem{Greub:2000sy}
C.~Greub and P.~Liniger, \emph{{Calculation of next-to-leading QCD corrections
  to b ---\ensuremath{>} sg}},
  \href{https://doi.org/10.1103/PhysRevD.63.054025}{\emph{Phys. Rev. D}
  {\bfseries 63} (2001) 054025}
  [\href{https://arxiv.org/abs/hep-ph/0009144}{{\ttfamily hep-ph/0009144}}].

\bibitem{Greub:2000an}
C.~Greub and P.~Liniger, \emph{{The Rare decay b ---\ensuremath{>} s gluon
  beyond leading logarithms}},
  \href{https://doi.org/10.1016/S0370-2693(00)01205-3}{\emph{Phys. Lett. B}
  {\bfseries 494} (2000) 237}
  [\href{https://arxiv.org/abs/hep-ph/0008071}{{\ttfamily hep-ph/0008071}}].

\bibitem{Krinner:2013cja}
F.~Krinner, A.~Lenz and T.~Rauh, \emph{{The inclusive decay $b \to c\bar{c}s$
  revisited}},
  \href{https://doi.org/10.1016/j.nuclphysb.2013.07.028}{\emph{Nucl. Phys. B}
  {\bfseries 876} (2013) 31} [\href{https://arxiv.org/abs/1305.5390}{{\ttfamily
  1305.5390}}].

\bibitem{Czarnecki:2005vr}
A.~Czarnecki, M.~Slusarczyk and F.~V. Tkachov, \emph{{Enhancement of the
  hadronic b quark decays}},
  \href{https://doi.org/10.1103/PhysRevLett.96.171803}{\emph{Phys. Rev. Lett.}
  {\bfseries 96} (2006) 171803}
  [\href{https://arxiv.org/abs/hep-ph/0511004}{{\ttfamily hep-ph/0511004}}].

\bibitem{Lenz:2022rbq}
A.~Lenz, M.~L. Piscopo and A.~V. Rusov, \emph{{Disintegration of beauty: a
  precision study}}, \href{https://doi.org/10.1007/JHEP01(2023)004}{\emph{JHEP}
  {\bfseries 01} (2023) 004}
  [\href{https://arxiv.org/abs/2208.02643}{{\ttfamily 2208.02643}}].

\bibitem{Beneke:2002rj}
M.~Beneke, G.~Buchalla, C.~Greub, A.~Lenz and U.~Nierste, \emph{{The $B^+
  -B^0_d$ Lifetime Difference Beyond Leading Logarithms}},
  \href{https://doi.org/10.1016/S0550-3213(02)00561-8}{\emph{Nucl. Phys. B}
  {\bfseries 639} (2002) 389}
  [\href{https://arxiv.org/abs/hep-ph/0202106}{{\ttfamily hep-ph/0202106}}].

\bibitem{Franco:2002fc}
E.~Franco, V.~Lubicz, F.~Mescia and C.~Tarantino, \emph{{Lifetime ratios of
  beauty hadrons at the next-to-leading order in QCD}},
  \href{https://doi.org/10.1016/S0550-3213(02)00262-6}{\emph{Nucl. Phys. B}
  {\bfseries 633} (2002) 212}
  [\href{https://arxiv.org/abs/hep-ph/0203089}{{\ttfamily hep-ph/0203089}}].

\bibitem{Gabbiani:2004tp}
F.~Gabbiani, A.~I. Onishchenko and A.~A. Petrov, \emph{{Spectator effects and
  lifetimes of heavy hadrons}},
  \href{https://doi.org/10.1103/PhysRevD.70.094031}{\emph{Phys. Rev. D}
  {\bfseries 70} (2004) 094031}
  [\href{https://arxiv.org/abs/hep-ph/0407004}{{\ttfamily hep-ph/0407004}}].

\bibitem{Gabbiani:2003pq}
F.~Gabbiani, A.~I. Onishchenko and A.~A. Petrov, \emph{{Lambda(b) lifetime
  puzzle in heavy quark expansion}},
  \href{https://doi.org/10.1103/PhysRevD.68.114006}{\emph{Phys. Rev. D}
  {\bfseries 68} (2003) 114006}
  [\href{https://arxiv.org/abs/hep-ph/0303235}{{\ttfamily hep-ph/0303235}}].

\bibitem{Lenz:2020oce}
A.~Lenz, M.~L. Piscopo and A.~V. Rusov, \emph{{Contribution of the Darwin
  operator to non-leptonic decays of heavy quarks}},
  \href{https://doi.org/10.1007/JHEP12(2020)199}{\emph{JHEP} {\bfseries 12}
  (2020) 199} [\href{https://arxiv.org/abs/2004.09527}{{\ttfamily
  2004.09527}}].

\bibitem{Mannel:2023zei}
T.~Mannel, D.~Moreno and A.~A. Pivovarov, \emph{{Heavy-quark expansion for
  lifetimes: Toward the QCD corrections to power suppressed terms}},
  \href{https://doi.org/10.1103/PhysRevD.107.114026}{\emph{Phys. Rev. D}
  {\bfseries 107} (2023) 114026}
  [\href{https://arxiv.org/abs/2304.08964}{{\ttfamily 2304.08964}}].

\bibitem{in_prep}
M.~Egner, M.~Fael, A.~Lenz, M.~L. Piscopo, A.~Rusov, K.~Sch\"{o}nwald and
  M.~Steinhauser, ``in preparation.''

\bibitem{Gorbahn:2004my}
M.~Gorbahn and U.~Haisch, \emph{{Effective Hamiltonian for non-leptonic
  $|\Delta F| = 1$ decays at NNLO in QCD}},
  \href{https://doi.org/10.1016/j.nuclphysb.2005.01.047}{\emph{Nucl. Phys. B}
  {\bfseries 713} (2005) 291}
  [\href{https://arxiv.org/abs/hep-ph/0411071}{{\ttfamily hep-ph/0411071}}].

\bibitem{Chetyrkin:1997gb}
K.~G. Chetyrkin, M.~Misiak and M.~Munz, \emph{{$|\Delta F| = 1$ nonleptonic
  effective Hamiltonian in a simpler scheme}},
  \href{https://doi.org/10.1016/S0550-3213(98)00131-X}{\emph{Nucl. Phys. B}
  {\bfseries 520} (1998) 279}
  [\href{https://arxiv.org/abs/hep-ph/9711280}{{\ttfamily hep-ph/9711280}}].

\bibitem{Lenz:2022pgw}
A.~Lenz, J.~M\"uller, M.~L. Piscopo and A.~V. Rusov, \emph{{Taming new physics
  in b \textrightarrow{} c\={u}d(s) with
  \ensuremath{\tau}(B$^{+}$)/\ensuremath{\tau}(B$_{d}$) and $ {a}_{sl}^d $}},
  \href{https://doi.org/10.1007/JHEP09(2023)028}{\emph{JHEP} {\bfseries 09}
  (2023) 028} [\href{https://arxiv.org/abs/2211.02724}{{\ttfamily
  2211.02724}}].

\bibitem{Lenz:1998qp}
A.~Lenz, U.~Nierste and G.~Ostermaier, \emph{{Determination of the CKM angle
  gamma and |V(ub) / V(cb)| from inclusive direct CP asymmetries and branching
  ratios in charmless B decays}},
  \href{https://doi.org/10.1103/PhysRevD.59.034008}{\emph{Phys. Rev. D}
  {\bfseries 59} (1999) 034008}
  [\href{https://arxiv.org/abs/hep-ph/9802202}{{\ttfamily hep-ph/9802202}}].

\bibitem{Lenz:1997aa}
A.~Lenz, U.~Nierste and G.~Ostermaier, \emph{{Penguin diagrams, charmless B
  decays and the missing charm puzzle}},
  \href{https://doi.org/10.1103/PhysRevD.56.7228}{\emph{Phys. Rev. D}
  {\bfseries 56} (1997) 7228}
  [\href{https://arxiv.org/abs/hep-ph/9706501}{{\ttfamily hep-ph/9706501}}].

\bibitem{Nogueira:1991ex}
P.~Nogueira, \emph{{Automatic Feynman Graph Generation}},
  \href{https://doi.org/10.1006/jcph.1993.1074}{\emph{J. Comput. Phys.}
  {\bfseries 105} (1993) 279}.

\bibitem{Gerlach:2022qnc}
M.~Gerlach, F.~Herren and M.~Lang, \emph{{tapir: A tool for topologies,
  amplitudes, partial fraction decomposition and input for reductions}},
  \href{https://doi.org/10.1016/j.cpc.2022.108544}{\emph{Comput. Phys. Commun.}
  {\bfseries 282} (2023) 108544}
  [\href{https://arxiv.org/abs/2201.05618}{{\ttfamily 2201.05618}}].

\bibitem{Kuipers:2012rf}
J.~Kuipers, T.~Ueda, J.~A.~M. Vermaseren and J.~Vollinga, \emph{{FORM version
  4.0}}, \href{https://doi.org/10.1016/j.cpc.2012.12.028}{\emph{Comput. Phys.
  Commun.} {\bfseries 184} (2013) 1453}
  [\href{https://arxiv.org/abs/1203.6543}{{\ttfamily 1203.6543}}].

\bibitem{Harlander:1998cmq}
R.~Harlander, T.~Seidensticker and M.~Steinhauser, \emph{{Complete corrections
  of Order alpha alpha-s to the decay of the Z boson into bottom quarks}},
  \href{https://doi.org/10.1016/S0370-2693(98)00220-2}{\emph{Phys. Lett. B}
  {\bfseries 426} (1998) 125}
  [\href{https://arxiv.org/abs/hep-ph/9712228}{{\ttfamily hep-ph/9712228}}].

\bibitem{Seidensticker:1999bb}
T.~Seidensticker, \emph{{Automatic application of successive asymptotic
  expansions of Feynman diagrams}},  in \emph{{6th International Workshop on
  New Computing Techniques in Physics Research: Software Engineering,
  Artificial Intelligence Neural Nets, Genetic Algorithms, Symbolic Algebra,
  Automatic Calculation}}, 5, 1999,
  \href{https://arxiv.org/abs/hep-ph/9905298}{{\ttfamily hep-ph/9905298}}.

\bibitem{Klappert:2020nbg}
J.~Klappert, F.~Lange, P.~Maierh\"ofer and J.~Usovitsch, \emph{{Integral
  reduction with Kira 2.0 and finite field methods}},
  \href{https://doi.org/10.1016/j.cpc.2021.108024}{\emph{Comput. Phys. Commun.}
  {\bfseries 266} (2021) 108024}
  [\href{https://arxiv.org/abs/2008.06494}{{\ttfamily 2008.06494}}].

\bibitem{fermat}
R.~H. Lewis, ``Computer algebra system fermat.''
  \texttt{https://home.bway.net/lewis}.

\bibitem{Klappert:2019emp}
J.~Klappert and F.~Lange, \emph{{Reconstructing rational functions with
  FireFly}}, \href{https://doi.org/10.1016/j.cpc.2019.106951}{\emph{Comput.
  Phys. Commun.} {\bfseries 247} (2020) 106951}
  [\href{https://arxiv.org/abs/1904.00009}{{\ttfamily 1904.00009}}].

\bibitem{Klappert:2020aqs}
J.~Klappert, S.~Y. Klein and F.~Lange, \emph{{Interpolation of dense and sparse
  rational functions and other improvements in FireFly}},
  \href{https://doi.org/10.1016/j.cpc.2021.107968}{\emph{Comput. Phys. Commun.}
  {\bfseries 264} (2021) 107968}
  [\href{https://arxiv.org/abs/2004.01463}{{\ttfamily 2004.01463}}].

\bibitem{Smirnov:2020quc}
A.~V. Smirnov and V.~A. Smirnov, \emph{{How to choose master integrals}},
  \href{https://doi.org/10.1016/j.nuclphysb.2020.115213}{\emph{Nucl. Phys. B}
  {\bfseries 960} (2020) 115213}
  [\href{https://arxiv.org/abs/2002.08042}{{\ttfamily 2002.08042}}].

\bibitem{Meyer:2017joq}
C.~Meyer, \emph{{Algorithmic transformation of multi-loop master integrals to a
  canonical basis with CANONICA}},
  \href{https://doi.org/10.1016/j.cpc.2017.09.014}{\emph{Comput. Phys. Commun.}
  {\bfseries 222} (2018) 295}
  [\href{https://arxiv.org/abs/1705.06252}{{\ttfamily 1705.06252}}].

\bibitem{Lee:2020zfb}
R.~N. Lee, \emph{{Libra: A package for transformation of differential systems
  for multiloop integrals}},
  \href{https://doi.org/10.1016/j.cpc.2021.108058}{\emph{Comput. Phys. Commun.}
  {\bfseries 267} (2021) 108058}
  [\href{https://arxiv.org/abs/2012.00279}{{\ttfamily 2012.00279}}].

\bibitem{Lee:2014ioa}
R.~N. Lee, \emph{{Reducing differential equations for multiloop master
  integrals}}, \href{https://doi.org/10.1007/JHEP04(2015)108}{\emph{JHEP}
  {\bfseries 04} (2015) 108} [\href{https://arxiv.org/abs/1411.0911}{{\ttfamily
  1411.0911}}].

\bibitem{Henn:2013pwa}
J.~M. Henn, \emph{{Multiloop integrals in dimensional regularization made
  simple}}, \href{https://doi.org/10.1103/PhysRevLett.110.251601}{\emph{Phys.
  Rev. Lett.} {\bfseries 110} (2013) 251601}
  [\href{https://arxiv.org/abs/1304.1806}{{\ttfamily 1304.1806}}].

\bibitem{Liu:2017jxz}
X.~Liu, Y.-Q. Ma and C.-Y. Wang, \emph{{A Systematic and Efficient Method to
  Compute Multi-loop Master Integrals}},
  \href{https://doi.org/10.1016/j.physletb.2018.02.026}{\emph{Phys. Lett. B}
  {\bfseries 779} (2018) 353}
  [\href{https://arxiv.org/abs/1711.09572}{{\ttfamily 1711.09572}}].

\bibitem{Liu:2021wks}
X.~Liu and Y.-Q. Ma, \emph{{Multiloop corrections for collider processes using
  auxiliary mass flow}},
  \href{https://doi.org/10.1103/PhysRevD.105.L051503}{\emph{Phys. Rev. D}
  {\bfseries 105} (2022) L051503}
  [\href{https://arxiv.org/abs/2107.01864}{{\ttfamily 2107.01864}}].

\bibitem{Liu:2022chg}
X.~Liu and Y.-Q. Ma, \emph{{AMFlow: A Mathematica package for Feynman integrals
  computation via auxiliary mass flow}},
  \href{https://doi.org/10.1016/j.cpc.2022.108565}{\emph{Comput. Phys. Commun.}
  {\bfseries 283} (2023) 108565}
  [\href{https://arxiv.org/abs/2201.11669}{{\ttfamily 2201.11669}}].

\bibitem{pslq}
H.~R.~P. Ferguson, D.~H. Bailey and S.~Arno, \emph{Analysis of pslq, an integer
  relation finding algorithm}, {\emph{Mathematics of Computation} {\bfseries
  68} (1999) 351}.

\bibitem{Remiddi:1999ew}
E.~Remiddi and J.~A.~M. Vermaseren, \emph{{Harmonic polylogarithms}},
  \href{https://doi.org/10.1142/S0217751X00000367}{\emph{Int. J. Mod. Phys. A}
  {\bfseries 15} (2000) 725}
  [\href{https://arxiv.org/abs/hep-ph/9905237}{{\ttfamily hep-ph/9905237}}].

\bibitem{Ablinger:2011te}
J.~Ablinger, J.~Blumlein and C.~Schneider, \emph{{Harmonic Sums and
  Polylogarithms Generated by Cyclotomic Polynomials}},
  \href{https://doi.org/10.1063/1.3629472}{\emph{J. Math. Phys.} {\bfseries 52}
  (2011) 102301} [\href{https://arxiv.org/abs/1105.6063}{{\ttfamily
  1105.6063}}].

\bibitem{Goncharov:1998kja}
A.~B. Goncharov, \emph{{Multiple polylogarithms, cyclotomy and modular
  complexes}}, \href{https://doi.org/10.4310/MRL.1998.v5.n4.a7}{\emph{Math.
  Res. Lett.} {\bfseries 5} (1998) 497}
  [\href{https://arxiv.org/abs/1105.2076}{{\ttfamily 1105.2076}}].

\bibitem{Goncharov:2001iea}
A.~B. Goncharov, \emph{{Multiple polylogarithms and mixed Tate motives}},
  \href{https://arxiv.org/abs/math/0103059}{{\ttfamily math/0103059}}.

\bibitem{Fael:2021kyg}
M.~Fael, F.~Lange, K.~Sch\"onwald and M.~Steinhauser, \emph{{A semi-analytic
  method to compute Feynman integrals applied to four-loop corrections to the $
  \overline{\mathrm{MS}} $-pole quark mass relation}},
  \href{https://doi.org/10.1007/JHEP09(2021)152}{\emph{JHEP} {\bfseries 09}
  (2021) 152} [\href{https://arxiv.org/abs/2106.05296}{{\ttfamily
  2106.05296}}].

\bibitem{Fael:2022rgm}
M.~Fael, F.~Lange, K.~Sch\"onwald and M.~Steinhauser, \emph{{Massive Vector
  Form Factors to Three Loops}},
  \href{https://doi.org/10.1103/PhysRevLett.128.172003}{\emph{Phys. Rev. Lett.}
  {\bfseries 128} (2022) 172003}
  [\href{https://arxiv.org/abs/2202.05276}{{\ttfamily 2202.05276}}].

\bibitem{Fael:2022miw}
M.~Fael, F.~Lange, K.~Sch\"onwald and M.~Steinhauser, \emph{{Singlet and
  nonsinglet three-loop massive form factors}},
  \href{https://doi.org/10.1103/PhysRevD.106.034029}{\emph{Phys. Rev. D}
  {\bfseries 106} (2022) 034029}
  [\href{https://arxiv.org/abs/2207.00027}{{\ttfamily 2207.00027}}].

\bibitem{Fael:2023zqr}
M.~Fael, F.~Lange, K.~Sch\"onwald and M.~Steinhauser, \emph{{Massive three-loop
  form factors: Anomaly contribution}},
  \href{https://doi.org/10.1103/PhysRevD.107.094017}{\emph{Phys. Rev. D}
  {\bfseries 107} (2023) 094017}
  [\href{https://arxiv.org/abs/2302.00693}{{\ttfamily 2302.00693}}].

\bibitem{Egner:2023kxw}
M.~Egner, M.~Fael, K.~Sch\"onwald and M.~Steinhauser, \emph{{Revisiting
  semileptonic B meson decays at next-to-next-to-leading order}},
  \href{https://doi.org/10.1007/JHEP09(2023)112}{\emph{JHEP} {\bfseries 09}
  (2023) 112} [\href{https://arxiv.org/abs/2308.01346}{{\ttfamily
  2308.01346}}].

\bibitem{Davydychev:1999ic}
A.~I. Davydychev and V.~A. Smirnov, \emph{{Threshold expansion of the sunset
  diagram}}, \href{https://doi.org/10.1016/S0550-3213(99)00269-2}{\emph{Nucl.
  Phys. B} {\bfseries 554} (1999) 391}
  [\href{https://arxiv.org/abs/hep-ph/9903328}{{\ttfamily hep-ph/9903328}}].

\bibitem{Dugan:1990df}
M.~J. Dugan and B.~Grinstein, \emph{{On the vanishing of evanescent
  operators}}, \href{https://doi.org/10.1016/0370-2693(91)90680-O}{\emph{Phys.
  Lett. B} {\bfseries 256} (1991) 239}.

\bibitem{Bobeth:1999mk}
C.~Bobeth, M.~Misiak and J.~Urban, \emph{{Photonic penguins at two loops and
  $m_t$ dependence of $BR[B \to X_s l^+ l^-]$}},
  \href{https://doi.org/10.1016/S0550-3213(00)00007-9}{\emph{Nucl. Phys. B}
  {\bfseries 574} (2000) 291}
  [\href{https://arxiv.org/abs/hep-ph/9910220}{{\ttfamily hep-ph/9910220}}].

\bibitem{Buras:1991jm}
A.~J. Buras, M.~Jamin, M.~E. Lautenbacher and P.~H. Weisz, \emph{{Effective
  Hamiltonians for $\Delta S = 1$ and $\Delta B = 1$ nonleptonic decays beyond
  the leading logarithmic approximation}},
  \href{https://doi.org/10.1016/0550-3213(92)90345-C}{\emph{Nucl. Phys. B}
  {\bfseries 370} (1992) 69}.

\bibitem{Chetyrkin:2000yt}
K.~G. Chetyrkin, J.~H. Kuhn and M.~Steinhauser, \emph{{RunDec: A Mathematica
  package for running and decoupling of the strong coupling and quark masses}},
  \href{https://doi.org/10.1016/S0010-4655(00)00155-7}{\emph{Comput. Phys.
  Commun.} {\bfseries 133} (2000) 43}
  [\href{https://arxiv.org/abs/hep-ph/0004189}{{\ttfamily hep-ph/0004189}}].

\bibitem{Herren:2017osy}
F.~Herren and M.~Steinhauser, \emph{{Version 3 of RunDec and CRunDec}},
  \href{https://doi.org/10.1016/j.cpc.2017.11.014}{\emph{Comput. Phys. Commun.}
  {\bfseries 224} (2018) 333}
  [\href{https://arxiv.org/abs/1703.03751}{{\ttfamily 1703.03751}}].

\bibitem{progdata}
https://www.ttp.kit.edu/preprints/2024/ttp24-020/.

\bibitem{egner_2024_11639756}
M.~Egner, M.~Fael, K.~Schönwald and M.~Steinhauser, ``{Supplemental material
  for ``Nonleptonic B-meson decays to next-to-next-to-leading order''}.'' URL:
  \url{https://doi.org/10.5281/zenodo.11639756}, 2024.

\bibitem{Voloshin:1994sn}
M.~B. Voloshin, \emph{{QCD radiative enhancement of the decay b
  ---\ensuremath{>} c anti-c s}},
  \href{https://doi.org/10.1103/PhysRevD.51.3948}{\emph{Phys. Rev. D}
  {\bfseries 51} (1995) 3948}
  [\href{https://arxiv.org/abs/hep-ph/9409391}{{\ttfamily hep-ph/9409391}}].

\bibitem{PDG}
S.~N. et~al. (Particle Data~Group). to be published in Phys. Rev. D 110, 030001
  (2024).

\bibitem{Buras:1992tc}
A.~J. Buras, M.~Jamin, M.~E. Lautenbacher and P.~H. Weisz, \emph{{Two loop
  anomalous dimension matrix for $\Delta S = 1$ weak nonleptonic decays I:
  $\mathcal{O}(\alpha_s^2)$}},
  \href{https://doi.org/10.1016/0550-3213(93)90397-8}{\emph{Nucl. Phys. B}
  {\bfseries 400} (1993) 37}
  [\href{https://arxiv.org/abs/hep-ph/9211304}{{\ttfamily hep-ph/9211304}}].

\bibitem{Ciuchini:1993vr}
M.~Ciuchini, E.~Franco, G.~Martinelli and L.~Reina, \emph{{The Delta S = 1
  effective Hamiltonian including next-to-leading order QCD and QED
  corrections}},
  \href{https://doi.org/10.1016/0550-3213(94)90118-X}{\emph{Nucl. Phys. B}
  {\bfseries 415} (1994) 403}
  [\href{https://arxiv.org/abs/hep-ph/9304257}{{\ttfamily hep-ph/9304257}}].

\bibitem{Chetyrkin:1996vx}
K.~G. Chetyrkin, M.~Misiak and M.~Munz, \emph{{Weak radiative B meson decay
  beyond leading logarithms}},
  \href{https://doi.org/10.1016/S0370-2693(97)00324-9}{\emph{Phys. Lett. B}
  {\bfseries 400} (1997) 206}
  [\href{https://arxiv.org/abs/hep-ph/9612313}{{\ttfamily hep-ph/9612313}}].

\bibitem{Gambino:2003zm}
P.~Gambino, M.~Gorbahn and U.~Haisch, \emph{{Anomalous dimension matrix for
  radiative and rare semileptonic B decays up to three loops}},
  \href{https://doi.org/10.1016/j.nuclphysb.2003.09.024}{\emph{Nucl. Phys. B}
  {\bfseries 673} (2003) 238}
  [\href{https://arxiv.org/abs/hep-ph/0306079}{{\ttfamily hep-ph/0306079}}].

\bibitem{Gorbahn:2005sa}
M.~Gorbahn, U.~Haisch and M.~Misiak, \emph{{Three-loop mixing of dipole
  operators}}, \href{https://doi.org/10.1103/PhysRevLett.95.102004}{\emph{Phys.
  Rev. Lett.} {\bfseries 95} (2005) 102004}
  [\href{https://arxiv.org/abs/hep-ph/0504194}{{\ttfamily hep-ph/0504194}}].

\bibitem{Czakon:2006ss}
M.~Czakon, U.~Haisch and M.~Misiak, \emph{{Four-Loop Anomalous Dimensions for
  Radiative Flavour-Changing Decays}},
  \href{https://doi.org/10.1088/1126-6708/2007/03/008}{\emph{JHEP} {\bfseries
  03} (2007) 008} [\href{https://arxiv.org/abs/hep-ph/0612329}{{\ttfamily
  hep-ph/0612329}}].

\bibitem{Gorbahn:2003zz}
M.~Gorbahn, \emph{{QCD and QED anomalous dimension matrix for weak decays at
  NNLO}},  other thesis, 10, 2003.

\bibitem{Herrlich:1994kh}
S.~Herrlich and U.~Nierste, \emph{{Evanescent operators, scheme dependences and
  double insertions}},
  \href{https://doi.org/10.1016/0550-3213(95)00474-7}{\emph{Nucl. Phys. B}
  {\bfseries 455} (1995) 39}
  [\href{https://arxiv.org/abs/hep-ph/9412375}{{\ttfamily hep-ph/9412375}}].

\bibitem{Misiak:1999yg}
M.~Misiak and J.~Urban, \emph{{QCD corrections to FCNC decays mediated by Z
  penguins and W boxes}},
  \href{https://doi.org/10.1016/S0370-2693(99)00150-1}{\emph{Phys. Lett. B}
  {\bfseries 451} (1999) 161}
  [\href{https://arxiv.org/abs/hep-ph/9901278}{{\ttfamily hep-ph/9901278}}].

\bibitem{Steinhauser:2000ry}
M.~Steinhauser, \emph{{MATAD: A Program package for the computation of MAssive
  TADpoles}},
  \href{https://doi.org/10.1016/S0010-4655(00)00204-6}{\emph{Comput. Phys.
  Commun.} {\bfseries 134} (2001) 335}
  [\href{https://arxiv.org/abs/hep-ph/0009029}{{\ttfamily hep-ph/0009029}}].

\bibitem{Buras:1998raa}
A.~J. Buras, \emph{{Weak Hamiltonian, CP violation and rare decays}},  in
  \emph{{Les Houches Summer School in Theoretical Physics, Session 68: Probing
  the Standard Model of Particle Interactions}}, pp.~281--539, 6, 1998,
  \href{https://arxiv.org/abs/hep-ph/9806471}{{\ttfamily hep-ph/9806471}}.

\bibitem{Buras:2006gb}
A.~J. Buras, M.~Gorbahn, U.~Haisch and U.~Nierste, \emph{{Charm quark
  contribution to K+ ---\ensuremath{>} pi+ nu anti-nu at
  next-to-next-to-leading order}},
  \href{https://doi.org/10.1007/JHEP11(2012)167}{\emph{JHEP} {\bfseries 11}
  (2006) 002} [\href{https://arxiv.org/abs/hep-ph/0603079}{{\ttfamily
  hep-ph/0603079}}].

\end{thebibliography}\endgroup

\end{document}